\documentclass[amsmath,amssymb]{revtex4}

\usepackage{color}

\usepackage{graphicx}
\usepackage{dcolumn}
\usepackage{bm}
\usepackage{epstopdf}
\usepackage{float}
\usepackage{amsmath}
\usepackage{soul}
\usepackage{array}
\usepackage{booktabs}
\input epsf
\usepackage{caption}
\usepackage{subfigure}
\usepackage{multirow}

\newcommand\beq{\begin{equation}}
\newcommand\eeq{\end{equation}}

\usepackage{color} 
\usepackage[english]{babel}
\usepackage[autostyle]{csquotes}

\begin{document}
    \title{Deformation and breakup of droplets in an oblique continuous air stream}
    \author{Surendra Kumar Soni$^{\dagger}$}
    \author{Pavan Kumar Kirar$^{\dagger\dagger}$}
    \author{Pankaj Kolhe$^{\dagger *}$}
    \author{Kirti Chandra Sahu$^{\dagger\dagger}$}\email{ksahu@iith.ac.in;psk@iith.ac.in}
    \affiliation{
        $^{\dagger}$Department of Mechanical and Aerospace Engineering, Indian Institute of Technology Hyderabad, Sangareddy 502 285, Telangana, India\\
        $^{\dagger\dagger}$Department of Chemical Engineering, Indian Institute of Technology Hyderabad, Sangareddy 502 285, Telangana, India}

 \begin{abstract}
We experimentally investigate the deformation and breakup of droplets interacting with an oblique continuous air stream. A high-speed imaging system is employed to record the trajectories and topological changes of the droplets of different liquids. The droplet size, the orientation of the air nozzle to the horizontal and fluid properties (surface tension and viscosity) are varied to study different breakup modes. We found that droplet possessing initial momentum prior to entering the continuous air stream exhibits a variation in the required Weber number for the vibrational to the bag breakup transition with a change in the angle of the air stream. The critical Weber numbers $(We_{cr})$ for the bag-type breakup are obtained as a function of the E\"{o}tv\"{o}s number $(Eo)$, angle of inclination of the air stream $(\alpha)$ and the Ohnesorge number $(Oh)$. It is found that although the droplet follows a rectilinear motion initially that transforms to a curvilinear motion at later times when the droplet undergoes topological changes. The apparent acceleration of the droplet and its size influence the critical Weber number for the bag breakup mode. The departure from the cross-flow arrangement shows a sharp decrease in the critical Weber number for the bag breakup which asymptotically reaches to a value associated with the in-line (opposed) flow configuration for the droplet breakup. 
\end{abstract}
    
\maketitle
    
\section{Introduction}
\label{sec:intro}
Fragmentation of liquid droplets due to the flow of high-speed air stream is observed in many industrial applications \cite{sikroria2014experimental} and natural phenomena \cite{Villermaux2009}. In atomization, initially fuel jet forms ligaments and breaks into droplets (primary atomization). These droplets become unstable when the aerodynamic force overcomes the influence of the surface tension force and the viscous force, and subsequently break into smaller satellite droplets (secondary atomization). The surface area to volume ratio significantly increases in the secondary atomization process, which in turn increases its efficiency (see e.g. Refs. \cite{varga2003initial,lefebvre2017atomization}). Thus fragmentation of a liquid droplet into tiny satellite droplets in in-line and cross-flow configurations has been a subject of research for the past several decades (see for instance \cite{taylor1963shape,Villermaux2009,Jain2015,dai2001temporal,pilch1987,chou1998temporal,krzeczkowski1980measurement}). 

Previous experiments suggest that there exists different types of mode when a liquid droplet interacts with high speed air stream, namely, the vibration, bag, bag-stamen, dual-bag, multi-mode, shear and catastrophic breakup modes \cite{dai2001temporal,guildenbecher2009,cao2007,Suryaprakash2019} owing to different mechanisms and flow characteristics. The Weber number that measures of the relative importance of the inertia over the surface tension force is mainly used to demonstrate different breakup modes. The Weber number is given by $We \equiv {\rho_a U^2 D / \sigma}$. The other dimensionless numbers which have been used to study droplet breakup phenomenon are the Reynolds number, $Re \equiv {\rho_l U D / \mu_l}$ (inertia force/viscous force), the E\"{o}tv\"{o}s number $Eo \equiv {(\rho_l-\rho_a) g D^{2} / \sigma}$ (gravitational force/surface tension force) and the Ohnesorge number, $Oh \equiv \sqrt{We}/ Re = {\mu_{l} / \sqrt{\rho_l \sigma D}}$, wherein, $\sigma$ is the interfacial tension, $g$ is the acceleration due to gravity, $D$ is the diameter of the droplet, $\rho_a$ and $\rho_l$ are the density of the air and liquid, respectively, and $U$ is the average velocity of the air stream. 

At low Weber numbers, a droplet undergoes shape oscillations at a certain frequency (vibrational regime). As the Weber number is increased slowly by increasing the aerodynamic force, and keeping the surface tension force constant, the droplet exhibits a transition from the vibrational mode to the bag breakup mode. The value of the Weber number at which this transition occurs (i.e. the Weber number at which the droplet just starts to form a bag) is termed as the critical Weber number ($We_{cr}$). Taylor \cite{taylor1963shape} was the first to predict that a liquid droplet undergoes breakup above a critical Weber number. Since then many researchers \cite{Jain2015,dai2001temporal,pilch1987,chou1998temporal,krzeczkowski1980measurement} have reported different values of the critical Weber number in different flow configurations. In cross-flow configuration, Table \ref{tab2} presents the range of the Weber numbers associated with the bag breakup region obtained by previous researchers. The discrepancies in obtaining the critical Weber number can be attributed to the definition of the breakup regime and the different experimental facilities (shock tubes, wind tunnels, convergent nozzles, etc.) used in their investigations. Moreover, the fragmentation process, due to its inherent nature, is highly sensitive to the flow conditions. Among all the fragmentation regimes described above, the bag breakup regime has the widest range of applications since it occurs at relatively low Weber numbers (see for instance, Refs. \cite{dai2001temporal,guildenbecher2009,Jain2015,kulkarni2014,flock2012,cao2007,Villermaux2009}). The bag breakup occurs at the beginning of the secondary atomization process and shows similar characteristics with that of the bag-stamen, the multi-bag and the shear stripping breakup modes \cite{Jain2015}. Hanson {\it et al.} \cite{hanson1963} and Wierzba \cite{wierzba1990deformation} demonstrated that the critical Weber number decreases as the droplet size increases in cross-flow configuration. Their results are summarised in Table \ref{tab3}. 

 \begin{table}
        \caption{The Weber number range for the bag type breakup.} 
        \label{tab2} 
        \centering
        \begin{tabular}{|c|c|}
                \hline
               References       &             Weber number range               \\   \hline
              Pilch and Erdman \cite{pilch1987}      & $12 <  We \leq 50$ \\   \hline
               Guildenbecher {\it et al.} \cite{guildenbecher2009}    &  $11 <  We < 35$ \\   \hline 
               Krzeczkowski \cite{krzeczkowski1980}    &   $10 <  We <18$   \\   \hline
                Jain {\it et al.} \cite{Jain2015}      &   $12 <  We < 24$ \\   \hline 
               Hsiang and Faeth \cite{hsiang1993}      &   $We \leq 11 \pm 2$  \\ \hline 
              Wierzba \cite{wierzba1990deformation}       &   $13.7 <  We <14.07$  \\  \hline
                Kulkarni and Sojka \cite{kulkarni2014}     &  $12 <  We <16$  \\   \hline 
                Wang {\it et al.} \cite{wang2014}    &  $10 <  We < 35$    \\  \hline
        \end{tabular}
\end{table}

Three types approaches have been used to study the droplet breakup phenomena: (i) the shock tube approach \cite{hsiang1993,dai2001,chou1998,krzeczkowski1980} (ii) the continuous air jet method \cite{zhao2016,Jain2015,kulkarni2014,flock2012,cao2007} and (iii) the droplet tower approach \cite{Villermaux2009}. In the present study, we implement a continuous air jet method. The main components of this method are an air nozzle that can produce a top-hat velocity profile and a compressed air measuring and controlling device. The results obtained using the first and the second approaches can be made similar by ensuring that the droplet interacts in the continuous air stream. This is possible when the time taken by the droplet to cross the shear boundary layer is much smaller than the resident time of the droplet when the droplet deforms and breaks \cite{guildenbecher2009}. 
   
\begin{table}[H]
    \caption{The critical Weber number for different droplet sizes reported by Hanson {\it et al.} \cite{hanson1963} and Wierzba \cite{wierzba1990deformation}.}
    \label{tab3}
    \centering
    \begin{tabular}{|c|c|c|c|}
        \hline
       References   &    Fluids  &  $D$ (mm) & $We_{cr}$  \\ \hline
        \multirow{19}{*}{} & \multirow{4}{*}{} &  0.426 & 6.62 \\ \cline{3-4} 
        &                   &  0.285 & 7.18  \\ \cline{3-4} 
        &     Silicone oil (10 cSt)               &  0.180 & 7.93   \\ \cline{3-4} 
        &                   &  0.117 & 7.94 \\ \cline{2-4} 
        & \multirow{4}{*}{} &  0.531 & 10.4  \\ \cline{3-4} 
        &                   &  0.273 & 11.8a  \\ \cline{3-4} 
        &      Silicone oil (50 cSt)              &  0.210 & 13.5  \\ \cline{3-4} 
        &                   &   0.126 & 15.4   \\ \cline{2-4} 
        & \multirow{6}{*}{} &  0.540 & 13.1  \\ \cline{3-4} 
        Hanson {\it et al.}  \cite{hanson1963}    &                   &  0.338 & 14.3  \\ \cline{3-4} 
        &                   &   0.239 & 15.5 \\ \cline{3-4} 
        &                   &  0.25 & 16.2 \\ \cline{3-4} 
        &    Silicone oil (100 cSt)                & 0.185 & 22.6  \\ \cline{3-4} 
        &                   &  0.150 & 23.8 \\ \cline{2-4} 
        & \multirow{3}{*}{} &  0.391 &4.79  \\ \cline{3-4} 
        &                   & 0.180 & 6.37  \\ \cline{3-4} 
        &    Water          &   0.0945 & 7.14 \\ \cline{2-4} 
        & \multirow{2}{*}{} & 0.471 & 6.76  \\ \cline{3-4} 
        &     Methyl alcohol              & 0.186 & 7.45   \\ \hline
        Wierzba \cite{wierzba1990deformation}&    Water                   &     2.22-3.9 & 13.7 to 14.07 \\ \hline
    \end{tabular}
\end{table}

As the above review indicates all the previous studies considered small droplets either in the cross-flow (e.g. \cite{wierzba1990deformation,hanson1963} or in the in-line (e.g. \cite{ Inamura2009,Villermaux2009}) configurations. However, in many situations, such as swirling spray applications and during rain, droplets may interact with the continuous air stream at an angle. In such situations, apart from the aerodynamic, the viscous and the surface tension forces, the gravitational force also influences the droplet breakup morphology when the droplet size is comparable or larger than the capillary length scale, given by $l_c = \sqrt{\sigma/(\rho_l - \rho_a) g}$ \cite{brenner1993}. When $D \ge l_c$, the surface tension force will not be able to hold the drop in a spherical shape and the droplet will deform due to the effect of gravity, and thereby changing the critical force requirement for different types of breakup. To the best of our knowledge, such a study has not been conducted yet.

In the present work, we experimentally investigate the deformation and breakup of a droplet interacting with an air stream flowing at an angle, $\alpha$ to the horizontal direction. A high-speed imaging system is employed to record the trajectory and topological change of the droplet and the effect of obliquity of the air stream on the droplet breakup process has been investigated. The droplet size, the orientation of the air nozzle, fluid properties (surface tension and viscosity) are varied to study different breakup modes. The critical Weber number has been obtained for different values of the E\"{o}tv\"{o}s number $(Eo)$, the angle of inclination of the air stream $(\alpha)$ and the Ohnesorge number $(Oh)$. It is found that, although the droplet exhibits rectilinear motion during its entry into the oblique air stream and prior to the formation of the bag breakup mode, a curvilinear motion is depicted by the droplet at later times, which indicates the variation in forces acting on the droplet. The apparent acceleration experienced by the droplet and its size are linked with the variation in the critical Weber number requirement for the bag breakup mode. The departure from the cross-flow arrangement shows a sharp decrease in the critical Weber number for the bag breakup with asymptotic behaviour at higher oblique angles.  At high obliquity, the critical Weber number approaches a value of 6 as reported in the literature (see for instance, Ref. \cite{Villermaux2009}) for the inline (opposed) flow configuration for the droplet breakup. It is also noted that in the present study, the effort has been made to give a detailed description of the experimental procedure, the cautions taken during the experiments, the uncertainty analysis that confirms the repeatability of the experimental results.

The rest of the paper is organized as follows. A brief description of the mathematical analysis is given in Section \ref{sec:math}, the experimental procedure is discussed in Section \ref{sec:expt}, which is followed by the discussion of our experimental results in Section \ref{sec:dis}. Finally, concluding remarks are given in Section \ref{sec:conc}.

 \section{Mathematical formulation}
\label{sec:math}
\begin{figure}[h]
\centering
\includegraphics[width=0.45\textwidth]{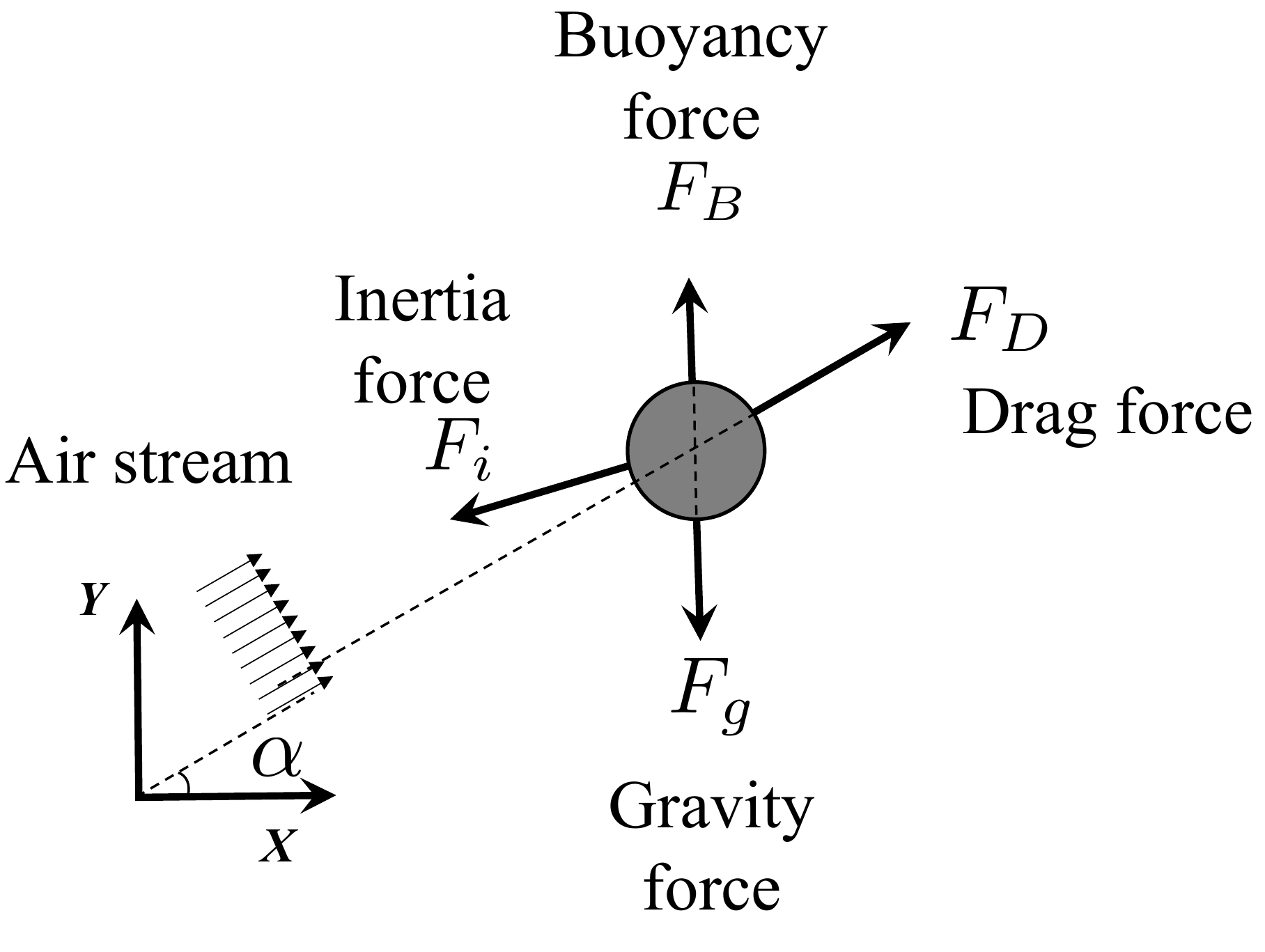} 
\caption{Schematic diagram showing a liquid droplet freely falling under the action of gravity and subjected to an air stream at an angle $\alpha$ to the horizontal.}
\label{fig1a}
\end{figure}

A schematic diagram showing a liquid droplet freely falling under the action of gravity and subjected to an air stream at an angle $\alpha$ to the horizontal is shown in Fig. \ref{fig1a}. In the crossed flow configuration, $\alpha=0$, such that the air stream flows in the horizontal $(X)$ direction and the droplet falls in the vertical $(Y)$ direction. For any oblique angle, $\alpha$ of the air stream, the various forces acting on the droplet are the gravity force $(F_g = {\pi \over 6} D^{3}\rho_l g)$, the buoyant force $(F_B = {\pi \over 6} {D}^{3}\rho_a g)$, the inertia force $(F_i =  \frac{\pi}{6} {D}^{3}\rho_a  \frac{d \textbf{u}_d}{dt})$ and the drag force ($F_D = {1 \over 2}C_d  \rho_a U^2 A$), such that $F_i= F_g -F_b + F_D$. Here $\textbf{u}_d$ is the velocity of the droplet, $A$ is the projected area of the deformed droplet and $C_d$ is drag coefficient and $t$ is time.

In the absence of the air stream, balancing the gravity force, $F_g$ with the surface tension force, $F_s = \pi D \sigma$, one can get the following expression for maximum droplet size, $D_{c}$ for which the surface tension dominates the other forces acting on the droplet while maintaining its sphericity. 
\begin{equation}
D_c= \sqrt{\frac{6\sigma}{(\rho_l-\rho_a)  g}}.
\end{equation}

In case of an oblique air stream, the droplet follows a curvilinear motion and experiences an apparent weight due to the rate of change of its acceleration. In this case, the resultant acceleration acting on the droplet ($\widehat{a}$) is given by
\begin{eqnarray}
 \widehat{a}  = \frac{F_i}{ \frac{\pi}{6} D^{3} \rho_l} = {g}+\frac{{F_D}}{ \frac{\pi}{6} D^{3} \rho_l}  = \underbrace{\left [{g}+\frac{{F_D} ~sin{\alpha}}{ \frac{\pi}{6} D^{3} \rho_l}  \right ]}_{\rm {Vertical ~ component}} +\underbrace{\left [ \frac{{F_D} ~cos{\alpha}}{ \frac{\pi}{6} D^{3} \rho_l}  \right ]}_{\rm Horizontal ~ component} 
    \label{eq:neta}
\end{eqnarray}
The net resultant acceleration have two components: the vertical, $\left ({g}+\frac{{F_D} ~ sin{\alpha}}{ \frac{\pi}{6} D^{3} \rho_l}  \right )$ and the horizontal component, $\left ( \frac{{F_D} ~ cos{\alpha}}{ \frac{\pi}{6} D^{3} \rho_l}  \right )$. Thus, for the cross-flow condition $(\alpha=0^\circ)$, the the vertical and the horizontal components of the acceleration are ${g}$ and $\left ( \frac{{F_D} }{ \frac{\pi}{6} D^{3} \rho_l}  \right )$, respectively. For the in-line (counter) flow configuration ($\alpha =90^\circ$), the component of acceleration in the vertical direction is $\left( {g} + \frac{{F_D} }{ \frac{\pi}{6} D^{3} \rho_l} \right)$ and the droplet does not encounter any acceleration in the horizontal direction.
    
Using the apparent weight of the droplet, the modified critical droplet diameter to maintained its sphericity can be obtained as
\begin{equation}
\widehat{D_c} = \sqrt{\frac{6\sigma}{(\rho_l-\rho_a) \widehat{a}}}.
\end{equation}

Thus for a droplet with $D>\widehat{D_c}$, the surface tension is weaker than the external force and the droplet deforms from its spherical shape.

Now we like to turn the attention to the time scales used in the previous studies. Different researchers have used different time scales to non-dimensionalise their results \cite{krzeczkowski1980,jalaal2012fragmentation}. Most of the researchers non-dimensionalised the breakup time with the transport time (a ratio of droplet diameter to its velocity) as suggested by Nicholls and Ranger \cite{nicholls1969}. Such nondimensionalisation provides an opposite trend as compared to the actual observed breakup time in case of oblique flow configurations. Considering the droplet breakup process being dictated by forces acting on the droplet, we propose and use the following time scale, $t_s \equiv \rho_a U/(\rho_l - \rho_a) g$ to nondimensionalise time, in the present study. 

\section{Experimental set-up and procedure}
\label{sec:expt}

\begin{figure}[H]
\centering
\includegraphics[width=0.6\textwidth]{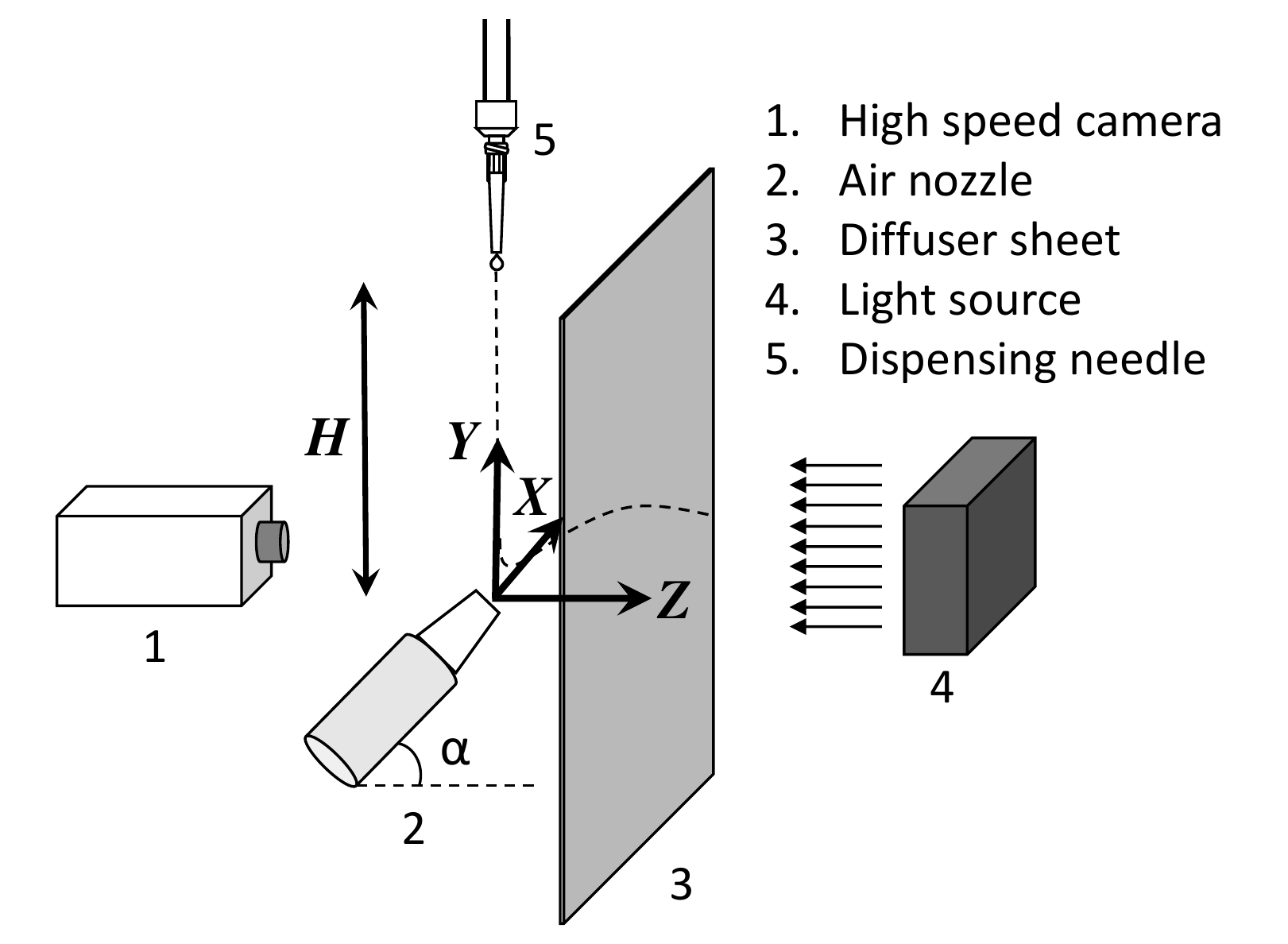}
\caption{Schematic of the experimental set-up. It consists of a high-speed camera, a nozzle, a diffuser sheet, a light source and a dispensing needle. The angle of inclination of the air nozzle to the horizontal is denoted by $\alpha$. $H$ ($\approx 40$ mm) is the height of the needle tip from the centre of the nozzle. Gravity, $g$ acts in the negative $Y$ direction. The orientation of the nozzle and the deflection of the nozzle is shown in $X-Y$ plane. }
\label{fig1}
\end{figure}

    Figure~\ref{fig1} depicts the schematic of the experimental set-up used in the present investigation. A metallic circular nozzle has been used to create a continuous and uniform air jet flow. This nozzle is hinged in such a way that the orientation of the nozzle can be fixed to any angle, $\alpha$ to the horizontal. A set of honeycomb pallet is placed near the upstream end of the nozzle to straighten the flow and to minimise disturbances. The internal section of the nozzle is designed to obtain a top-hat velocity profile at the nozzle exit, which has been verified with the aid of a pitot tube and differential head of the manometer measurements. The internal and external diameters of the pitot tube are 1.5 mm and 4 mm, respectively. The horizontal distance between the nozzle and the pitot tube is kept within 10 mm ($< 0.5$ diameter of the nozzle). For every experiment, it is ensured that the droplet enters in the continuous velocity profile of air or the potential jet core in the immediate downstream region of the nozzle exit before undergoing deformation and breakup. The air flow rate has been measured and controlled using an accurate and reliable ALICAT digital mass flow controller (range 0 to 2000 liter per minute) which is installed before the air nozzle.
    
    For generating the same size of liquid droplets, a regular infusion pump (SP-810 from Medical Point) is used. The drop size is varied in two ways: (i) by changing the mass flow rate setting in the infusion pump and (ii) by changing the spout having different exit diameters. The droplet generator and the circular nozzle are arranged in the cross-flow configuration ($\alpha=0^{\circ}$). The liquid droplet drips from the tip of the spout under the action of gravity and enters the air stream.
    
    A high-speed camera (Phantom VEO 710) with a 50 mm Nikon lens is used to capture the droplet deformation and breakup phenomena occurring near the nozzle. The images are recorded at 15000 frames per second (fps) with an exposure time of 10 $\mu$s for all the experiments conducted in the present study. The camera is connected to a computer and is operated via Phantom Camera Control (PCC) 3.1 software. The camera has a built-in memory card that allows for direct recording of the images. A diffused back-lit illumination (Light Emitting Diode (LED) based lighting, model 900445, 12000 lm, Visual Instrumentation Corporation) shadowgraph technique has been used to illuminate the background for droplet identification. In this technique, a light source is placed behind a diffusion screen to diffuse the backlit into uniform light. 
    
    The recorded images are analysed using the Matlab software. The geometric size of the droplet has been calculated by image processing. The droplet deforms while suspended in syringe spout and no longer remains perfectly spherical, hence the equivalent droplet initial diameter has been calculated as \cite{rioboo2002}
 \begin{equation}
D = (D^2_a \times D_b)^\frac{1}{3},
\end{equation}
where $D_a$ and $D_b$ are droplet diameter corresponding to the major axis and minor axis of the slightly deformed droplet.
    
The droplet velocity is found to be insignificant as compared to air velocity that allows us to use the absolute velocity of the air stream instead of the relative velocity in our calculations. Note that free jet is employed to investigate the droplet deformation and breakup. The uniform flow requirement for droplet breakup study is ensured through droplet entering the potential core region immediate downstream of the nozzle exit. For oblique air stream angles, if sufficient care is not taken, the droplet deformation and breakup process will occur in the shear layer instead of the jet potential core region. This will result in a higher critical Weber number as compared to the actual value. Thus, in the present study, sufficient care has been taken to ensure that droplet enters the potential core region of the air stream. 
    
In our experiments, the uniform size of droplets is created with the help of a syringe pump. For calculating the dimensionless numbers, the standard properties of the fluids are taken at $30^{\circ}$C. To see the effect of surface tension and viscosity on the droplet breakup phenomenon, four different fluids, namely, deionised water, 50 \%, 70 \% and 80 \% aqueous mixtures of glycerin (by volume) have been considered. The viscosities of the fluids are measured using a rotational rheometer (Model: RheolabQC). The density and surface tension of the fluids are taken from Ref. \cite{fp}. The density (kg/m$^3$), the surface tension (N/m) and the dynamic viscosity (mPa$\cdot$s) of the working fluids are given in Table~\ref{tab:4}.
    
    \begin{table}[ht]
        \caption{Physical properties of the working fluids.} 
        \label{tab:4} 
        \centering
       \begin{tabular}{|c|c|c|c|c|}
                \hline
                Working fluid  & Dynamic viscosity & Surface tension & Density  & Droplet diameter \\
                                & $\mu_l$ (mPa$\cdot$s) & $\sigma$ (mN/m) & $\rho_l$ (kg/m$^3$)  & $D$ (mm)  \\
                \hline
                Water               & 0.78  & 72 & 998 & 2.0-6.3 \\              \hline
                Glycerine (50\%)    & 4.85  & 67.5& 1127 & 2.5 -5.75 \\     \hline    
                Glycerine (70\%)    & 17.48  & 65.8 & 1182 & 2.8-5.75 \\     \hline
                Glycerine (80\%)    & 43.08  & 64.8 & 1209 & 3-5.6 \\     \hline    
        \end{tabular}
    \end{table}
    
    For current experiments, the nozzle orientation is varied from $\alpha=0^{\circ}$ to $\alpha=60^{\circ}$ with an interval of $10^{\circ}$  in order to investigate the droplet deformation and breakup.

    The uncertainty in the calculation of the field variables, like droplet diameter $(D)$, the air phase velocity and other dimensionless numbers are listed in Table~\ref{tab:5}. The maximum uncertainty propagation rule has been applied to calculate the uncertainty. According to this rule, if $p$, $q$, $r$, $...$ are the field variables having uncertainties $U_p$, $U_q$, $U_r$, $...$ in the measurements, the uncertainty in $Q = Q(p,q,r,...)$ can be calculated as follows
\begin{equation}
\delta Q = \sqrt{ \left ( \frac{\delta Q}{\delta p} \cdot U_{p}\right )^2     +  \left ( \frac{\delta Q}{\delta q} \cdot U_{q}\right )^2 +  \left ( \frac{\delta Q}{\delta r} \cdot U_{r}\right )^2 + .....} \quad . \label{uncer}
\end{equation}

    \begin{table}[ht]
        \caption{Uncertainty in the calculations of different variables in the present study.} 
        \label{tab:5} 
        \centering
        \begin{tabular}{|c|c|}
                \hline
                Variable &       Maximum uncertainty       \\  \hline
                $\alpha$ (degree) &        $5  \times 10^{-1}$        \\  \hline
                Air velocity (m/s) &     $6 \times 10^{-3}$         \\  \hline
                Weber number ($We$) &           $3.3 \times 10^{-1}$        \\   \hline
                E\"{o}tv\"{o}s number ($Eo$) &          $8 \times 10^{-2}$           \\   \hline
                Nozzle diameter (m) &         $5 \times 10^{-4}$      \\    \hline
                Droplet diameter (m)   &   $ 10^{-4}$ \\  \hline
        \end{tabular}
    \end{table}

\section{Results and discussion}
\label{sec:dis}

\begin{figure}[H]
\centering
\includegraphics[width=0.7\textwidth]{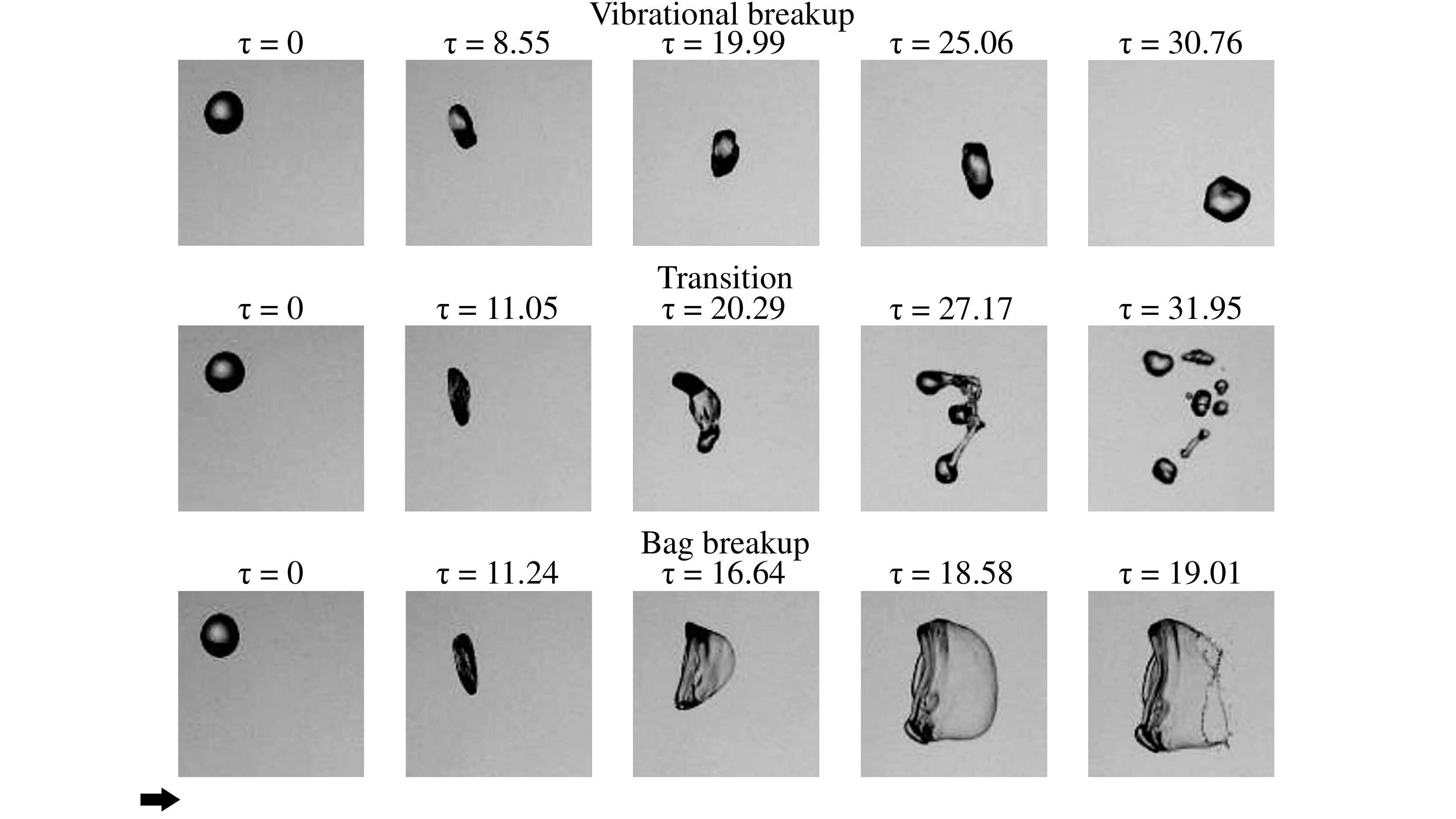}
\caption{The breakup dynamics of a water droplet in the cross-flow condition ($\alpha=0^\circ$) for $Eo \approx 2.4$ demonstrating the vibrational ($We=7.87$), the transitional ($We=8.27$) and the bag breakup ($We=9.13$) modes. The dimensionless times, $\tau$ are shown at the top each panel, such that $\tau=0$ represents the time when the droplet just enters the uniform air-stream.}
\label{fig2a}
\end{figure}

\begin{figure}[H]
\centering
\includegraphics[width=0.7\textwidth]{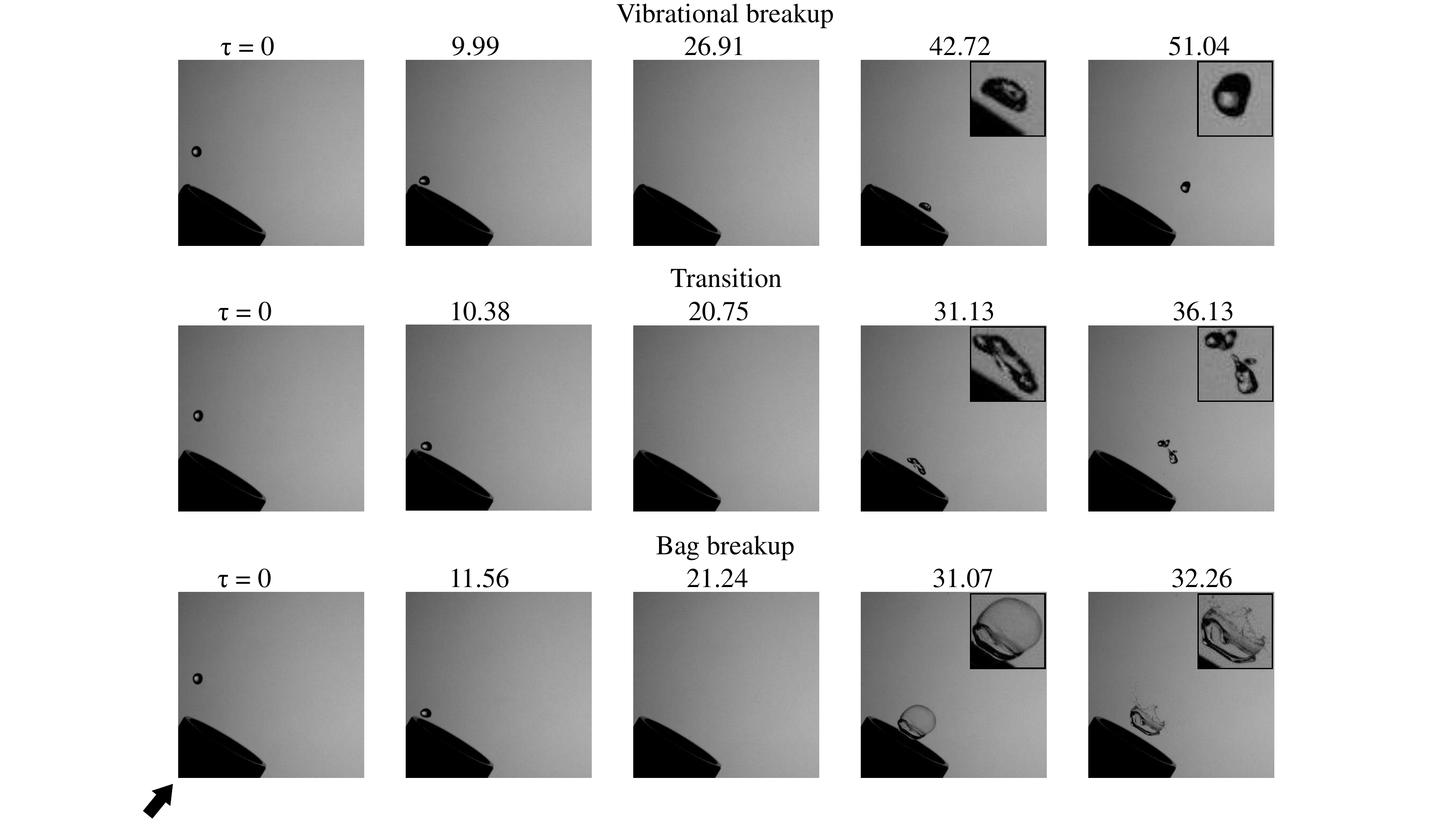}
\caption{The breakup dynamics of a water droplet in an oblique configuration ($\alpha=60^\circ$) for $Eo \approx 1.8$ demonstrating the vibrational ($We=5.13$), the transitional ($We=5.73$) and the bag breakup ($We=6.35$) modes. The dimensionless times, $\tau$ are shown at the top each panel, such that $\tau=0$ represents the time when the droplet just enters the uniform air-stream. The inset in each panel of the last two columns represent the zoomed view of the droplet.}
\label{fig2b}
\end{figure}

We begin the presentation of our results by demonstrating the topological changes of a freely falling water droplet under the action of gravity when subjected to an air stream flowing at an angle $\alpha=60^\circ$ to the horizontal. The dynamics is compared again the cross-flow configuration $(\alpha=0^\circ)$, which has been previously studied by several researchers (see for instance, Refs. \cite{guildenbecher2009,Jain2015,kulkarni2014,flock2012,cao2007}). Figs. \ref{fig2a} and \ref{fig2b} present the temporal variations of a droplet undergoing the vibrational, the transitional and the bag breakup for $\alpha=0^\circ$ and $60^\circ$, respectively. By varying the droplet size and air-stream velocity, $We$ and $Eo$ are varied to demonstrate the dynamics for $\alpha=0^\circ$ and $60^\circ$. The dimensionless time, $\tau \left( \equiv  (\rho_l-\rho_a) g t / U  \rho_a \right)$ is used to show the temporal evolutions of the droplet, such that $\tau=0$ represents the dimensionless time when the droplet just enters the uniform air-stream. 

For $\alpha=0^\circ$ and $Eo \approx 2.4$, it can be seen in the first row of Fig. \ref{fig2a} that (for $We=7.87$) the aerodynamic force is not enough to overcome the effect of the surface tension and viscous force, and the droplet undergoes shape oscillations (the vibrational regime). Due to the aerodynamic force, the amplitude of oscillations continue to increase and the droplet decomposes into comparable size droplets (not shown). In this case, it is not necessary that the disturbance created by the aerodynamic force is sufficient enough for the fragmentation process. The deformation time scale of the droplet undergoing the vibrational mode is also high as compared to the other modes. For a higher value of the Weber number ($We=8.27$), the droplet exhibits the transitional mode (second row of Fig. \ref{fig2a}) which is an intermediate stage between the vibrational mode and the bag breakup mode. As we increase the value of Weber number further ($We=9.13$) keeping the rest of the parameters fixed, the droplet exhibits the bag breakup mode as seen in the third row of Fig. \ref{fig2a}. The minimum value of the Weber number at which the droplet undergoes bag breakup phenomenon is known as the critical Weber number, $We_{cr}$. It can be concluded that for a fixed set of other parameters, the bag breakup mode can be achieved by increasing the air stream velocity only. 

The temporal evolutions for the vibrational, the transitional and the bag breakup modes for $\alpha=60^\circ$ and $Eo \approx 1.8$ are shown in Fig. \ref{fig2b} for $We=5.13$, 5.73 and 6.35, respectively. It can be seen that the phenomena are qualitatively similar to those observed in the cross-flow configuration. In all these cases, the droplet enters inside the air nozzle at $\tau \approx 10$, 10.38 and 11.56 and comes out at $\tau \approx 42.72$, 31.13 and 31.07 with oblate, disk and bag-type shapes for the vibrational, the transitional and the bag breakup modes, respectively. This confirms that even in the oblique configuration the droplet was interacting with the continuous air stream in the potential core region. 

\begin{figure}[H]
\centering
\hspace{0.5cm} {\large (a)} \hspace{7.4cm} {\large (b)}\\
\includegraphics[width=0.45\textwidth]{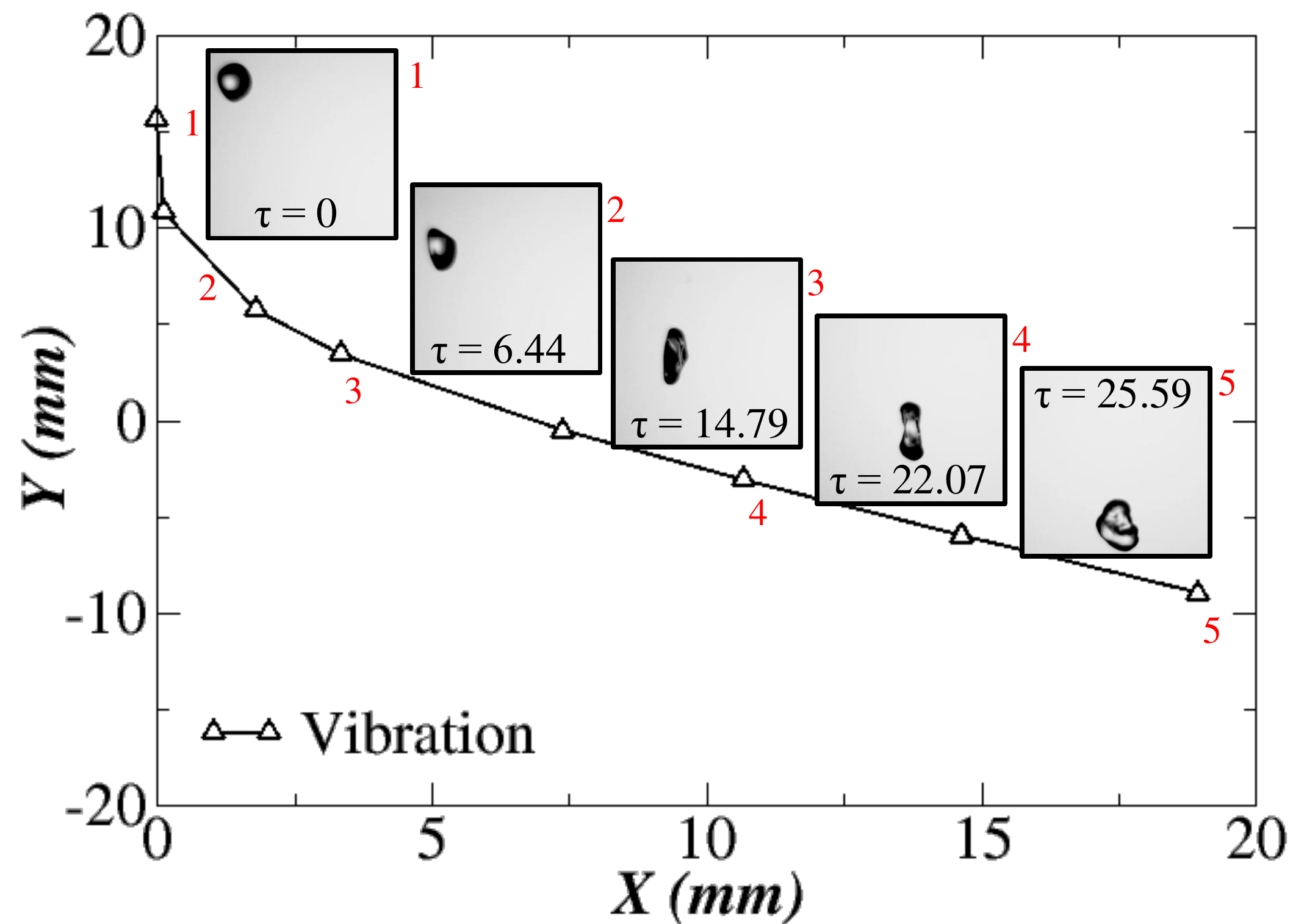}  \includegraphics[width=0.45\textwidth]{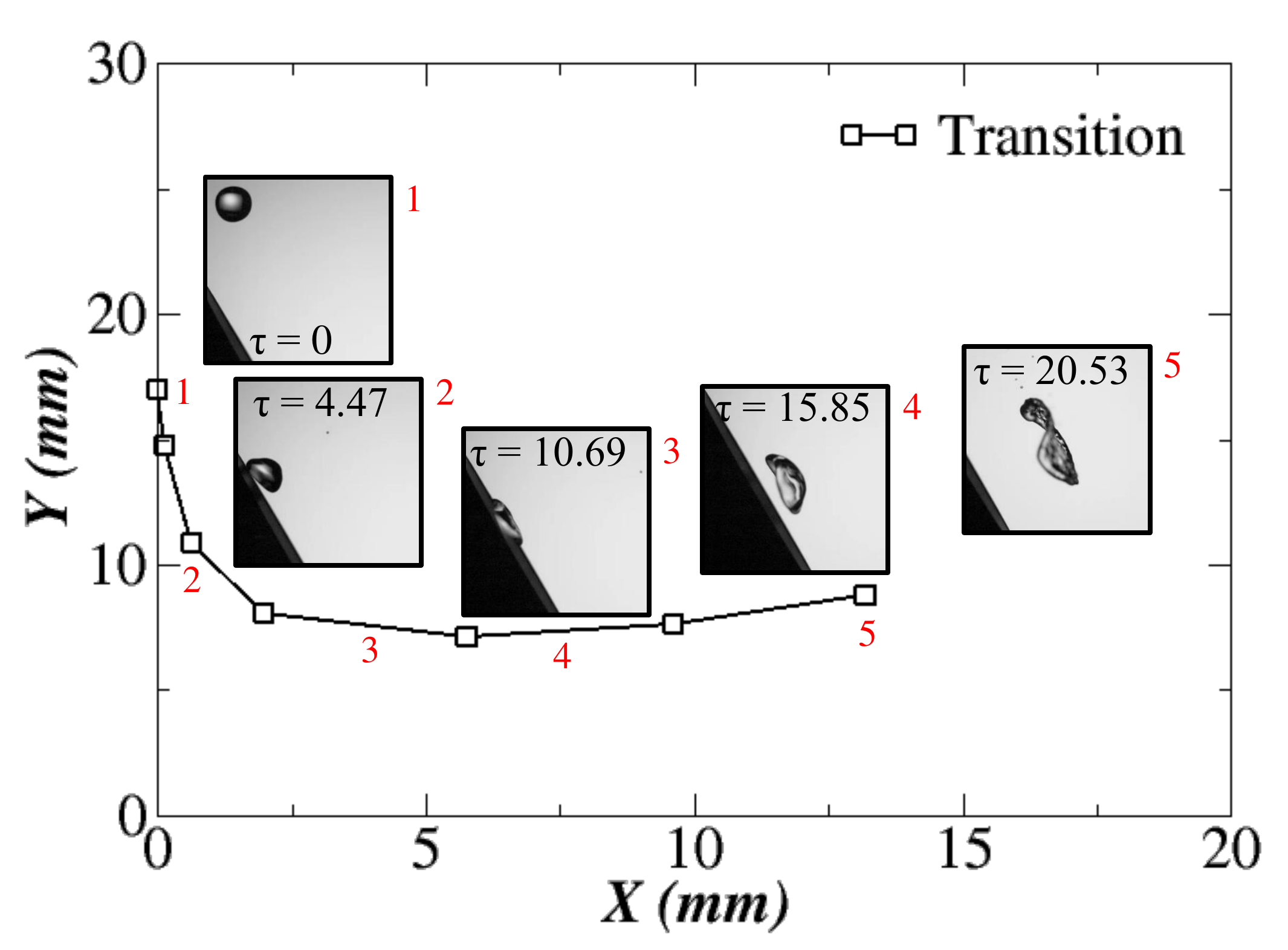} \\
\vspace{2mm}
\hspace{0.5cm} {\large (c)} \hspace{7.4cm} {\large (d)}\\
\includegraphics[width=0.45\textwidth]{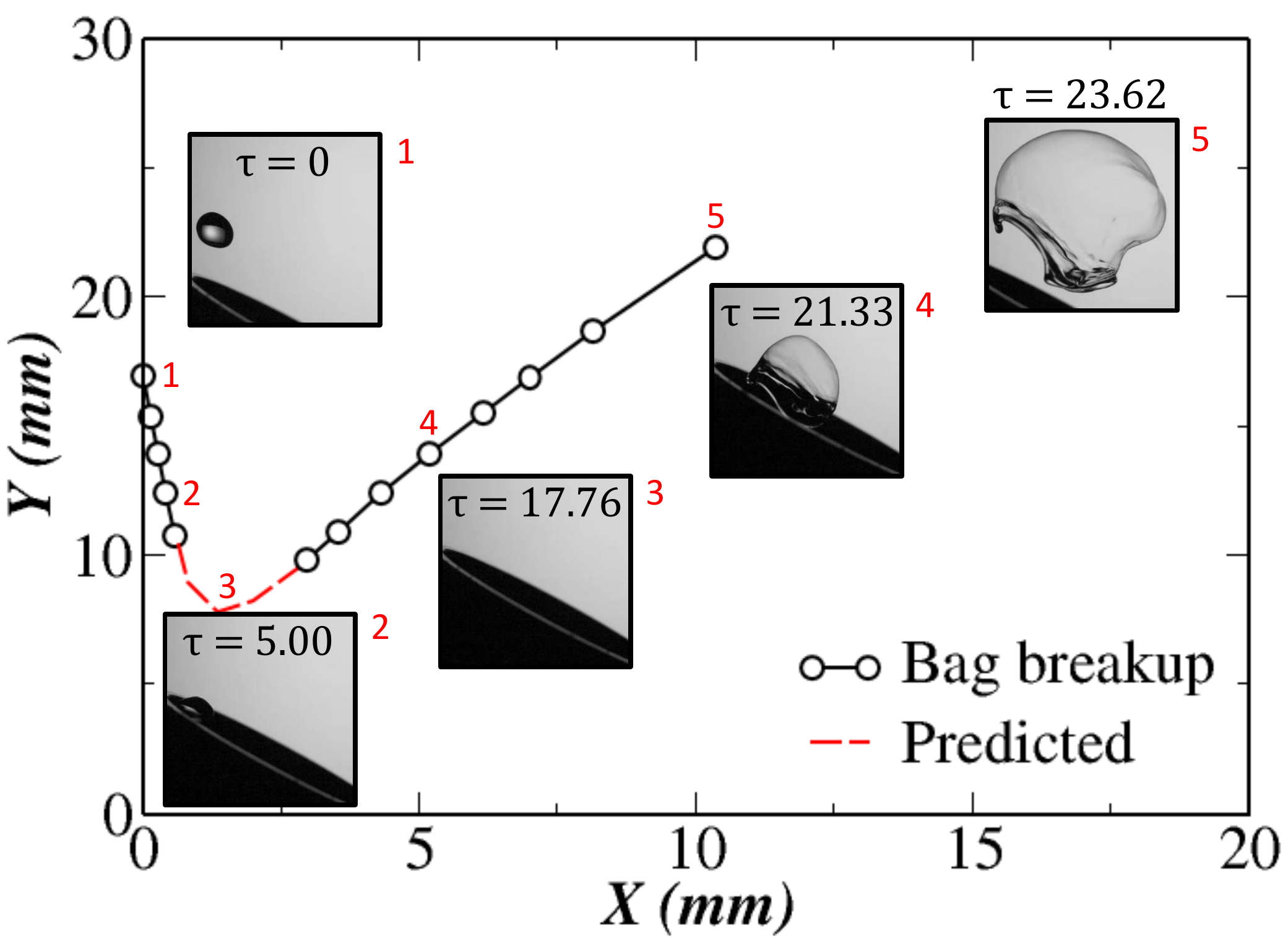} 
\includegraphics[width=0.45\textwidth]{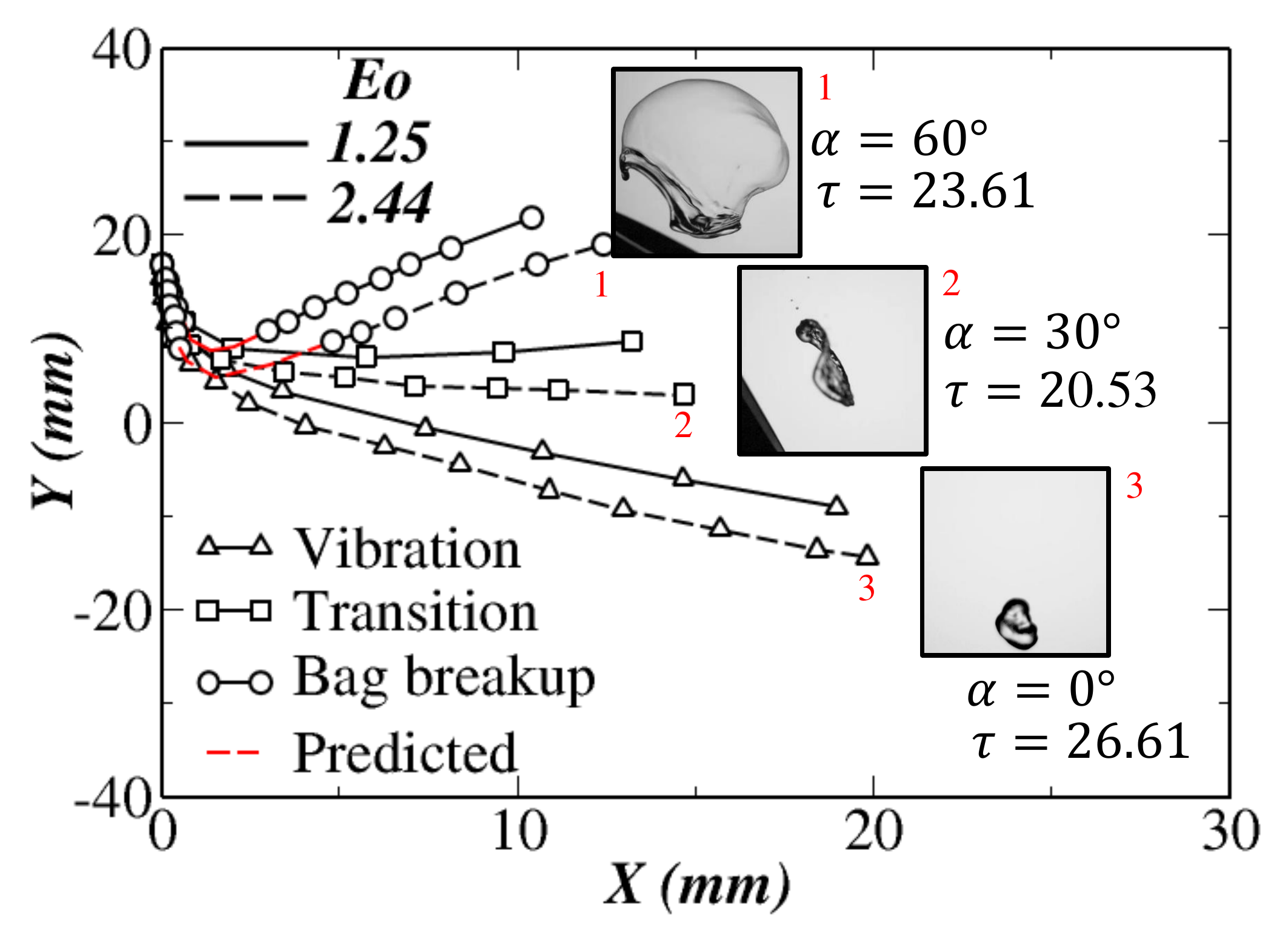} 
\caption{Trajectory and the shape of the droplet at different times for $Eo=1.25$ that corresponds to $D=3.03$ mm for different values of $\alpha$. (a) $\alpha=0^\circ$, (b)  $\alpha=30^\circ$ (c) $\alpha=60^\circ$. The panel (d) combines the trajectories for $Eo=1.25$ ($D=3.03$ mm) and $Eo=2.44$ ($D=4.23$ mm) at different oblique angles of the air stream. The numbers in each panel show the instants at which the droplet shapes are presented.}
 \label{fig:1a}
\end{figure}

Next, we investigate the effect of the air stream directionality on the droplet deformation and breakup phenomena. The drop size, height from the nozzle to tip, air flow velocity are kept constant and only the angle of the air stream, $\alpha$ is varied. Figs. \ref{fig:1a}(a), (b) and (c) present the temporal evolutions of the droplet along with their trajectories for $\alpha=0^\circ$, $30^\circ$ and $60^\circ$, respectively at $Eo=1.25$. Note that the Weber number can be calculated based on the relative velocity of the air stream with respect to the droplet or only the average velocity of the air stream from the nozzle. The parameters used to generate Fig. \ref{fig:1a} are given in Table~\ref{tab:1aa}. It can be seen that the droplet velocity, $u_d$ (calculated by the image processing method) is much smaller than the air stream velocity in all the cases. Two sizes of water droplet with $D=3.03$ mm and 4.23 mm are considered for which the values of the E\"{o}tv\"{o}s number are 1.25 and 2.44, respectively. It can be observed that the Weber number calculated based on the air stream velocity, $U$ is constant for each droplet size, but the Weber number calculated based on the relative velocity between the air stream and the droplet decreases slightly with the increase in the oblique angle of the air stream, $\alpha$. Thus, the Weber number, $We$ calculated based on the average air stream velocity is used to present the rest of the presented in this study. Note that in Fig. \ref{fig:1a}, only $\alpha$ is varied while the rest of the parameters remain fixed.

It can be seen in Fig. \ref{fig:1a}(a) for $\alpha=0^\circ$ that the droplet undergoes the vibrational mode. As we increase the angle of the air stream, the droplet exhibits a transitional mode for $\alpha=30^\circ$ (see Fig. \ref{fig:1a}(b)) and the bag breakup mode for $\alpha=60^\circ$ (see Fig. \ref{fig:1a}(c)). At $\tau$=0, the droplet is almost spherical for all the cases. In the vibrational mode, the non-uniform pressure distribution around the interface of the spherical drop initiates an internal flow within the drop, which in turn deforms the droplet to an oblate shape at $\tau \approx 6.44$ and a disk type shape at $\tau \approx 14.79$. In this case, the aerodynamic force is not enough to disintegrate the droplet into multiple pieces. Due to the competition between the aerodynamic force and the surface tension force, the droplet undergoes shape oscillations with a certain frequency. In some situations, these oscillations may lead to a breakup of the droplet into large ligament and satellite droplets of comparable sizes. 

\begin{table}[H]
\caption{The parameters considered to generate Fig. \ref{fig:1a}. Here, $We_r$ corresponds the Weber number defined based on the relative velocity between the air stream and the droplet.}  
\label{tab:1aa} 
\centering    
\begin{tabular}{|c|c|c|c|c|c|c|}
\hline
$\alpha$    & $D$ (mm) & $U$ (m/s)  & ${u}_d$ (m/s) & $Eo$ & $We$ & $We_r$    \\ \hline
$0^\circ$    & 3.03 & 12.33& 0.3 & 1.25 &  8.0   & 8.40   \\ 
& 4.23 & 9.82 & 0.3 & 2.44  & 7.1 &   7.54 \\  \hline
$30^\circ$    & 3.03 &12.33 & 0.3 & 1.25 & 8.0   &  8.34   \\
& 4.23 &9.82& 0.3 &2.44 & 7.1 &  7.48  \\ \hline
$60^\circ$    & 3.03 &12.33 & 0.3 &1.25 & 8.0   &  8.20  \\
& 4.23 &9.82 & 0.3 & 2.44 &   7.1&  7.32  \\ \hline
\end{tabular}
\end{table}   

For the same operating condition, as the air stream angle is increased to $\alpha$=30$^\circ$ the droplet undergoes a transitional region (Fig. \ref{fig:1a}(b)). In this case, as the droplet enters the air stream, the unequal distribution of the pressure changes the droplet shape from the spherical to a disk-like shape at $\tau \approx 10.69$. At $\tau \approx 15.85$, a bag-like structure appears in the air flow direction, but due to insufficient aerodynamic force, the droplet bag re-tracked back to a disk-like shape. Subsequent at later times $\tau > 20.53$, the droplet breaks into similar size droplets (not shown). The resultant satellite droplets exhibit vibrational modes.  

\begin{figure}[H]
    \centering
    \includegraphics[width=0.38\textwidth]{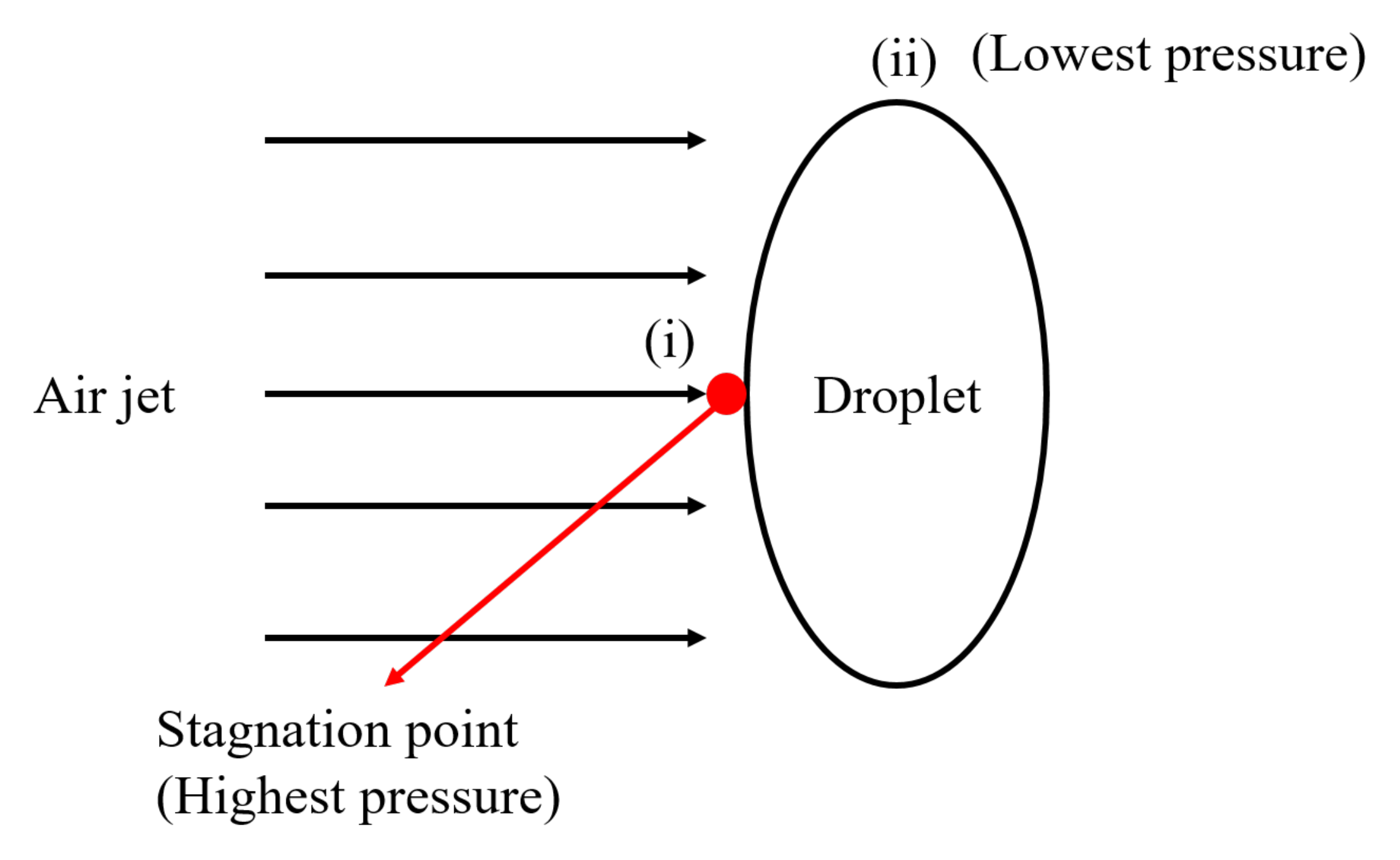}  
    \caption{Pressure distribution on the droplet subjected to air stream.}
    \label{fig:1b}
\end{figure}

At $\alpha=60^\circ$, the droplet clearly shows the bag-type breakup (Fig. \ref{fig:1a}(c)). In this case, when the spherical drop enters the region of the steady air stream, it is influenced by the distribution of air pressure around the drop. At an equilibrium condition, the internal and the surface tension forces are just balanced with the external aerodynamic force. However, when a steady air stream flows around a drop, the air velocity distribution and the air pressure distribution at any point on the drop surface are not uniform (see Fig. \ref{fig:1b}). The air velocity is maximum at point $ii$ and equals to zero at point $i$ (the stagnation point) in Fig. \ref{fig:1b}. Thus, by Bernoulli's equation, the air pressure becomes highest at point $i$, and lowest at point $ii$. In this situation, the external aerodynamic pressure causes the drop to deform to an oblate ellipsoid shape in the direction normal to the air flow. As the velocity at the equator of drop increases, the Bernoulli pressure difference increases accordingly, which flattens the drop more. Finally, the drop becomes a pancake-like shape (not shown as  the droplet is inside the air nozzle at this instant). Due to the contraction of the drop, a wake region \cite{flock2012} is developed on the back side of the droplet, which facilities the bag formation at $\tau \approx 19$. The bag carries more mass at the periphery due to the pressure distribution along the interface, that suppresses the instability while at the centre of the bag grows parallel to the air stream direction. The hollow portion of the bag becomes thin, like a membrane, and finally ruptures after $\tau \approx 23.62$ for this set of parameters. Due to this rupture, many satellite droplets are generated. Then the toroidal ring expands in the radial direction and breaks into relatively bigger child droplets in the downstream direction (not shown).

Comparison of the trajectories of the droplet in Figs. \ref{fig:1a}(a), (b) and (c) also reveals that as the oblique angle of the air stream, $\alpha$ increases, the droplet follow a curvilinear motion with a sharp turn and the distance travelled by droplet from the nozzle decreases. The change in projected area of the droplet changes the drag coefficient. As the tangent to the path of the droplet gives its velocity. The acceleration, thus the inertia force acting on the droplet, also changes in an oblique configuration. Thus, using such oblique configurations, one can design compact combustors with a small primary reaction zone. 

In Fig. \ref{fig:1a}(d), the trajectories of the droplets exhibiting the vibrational, the transitional and the bag breakup modes for $Eo=1.25$ ($D=3.03$ mm, solid lines) and 2.44 ($D=4.23$ mm, dashed lines) are shown. For both values of $Eo$, at $\alpha=0^\circ$, the droplet enters the aerodynamic field, deforms into a disk-like shape and dragged away in the direction of the air flow. The shape of the droplet for $\alpha=0^\circ$ is shown at $\tau$=25.59. At this time, the droplet is in the shear layer region outside the potential core region of the air stream. For higher values of $\alpha$, the droplet decelerates while migrating in the downward direction. When the droplet reaches the equilibrium stage, i.e., when the forces acting on the droplet in downward direction are balanced by the forces in the upward direction, the droplet comes to a stationary position for a very short time and starts to move in the opposite direction. At this stage, the droplet deforms to a shape that will facilitate it to migrate in the air flow direction. The droplet remains in contact with the strong aerodynamic force for a longer time as compared to the low $\alpha$ case (this residence time is proportional to the value of $\alpha$). This increases the probability to break the droplet even with smaller air velocity. Therefore, based on the above discussion, it can be concluded that the critical Weber number is sensitive to the air flow direction, $\alpha$. Increasing $\alpha$ decreases the critical Weber number for the bag breakup. This observation can play a significant role in the atomization process where the co and counter swirling secondary air streams are used to promote the secondary atomization process. The critical Weber number also sensitive to the droplet size. It is observed that for water droplets of diameters 3.03 $(Eo=1.25)$ and 4.23 mm ($Eo=2.44$), the values of $We_{cr}$ are 8 and 7.1, respectively.

\begin{figure}[H]
    \centering
\hspace{0.5cm} {\large (a)} \hspace{7.8cm} {\large (b)}\\
    \includegraphics[width=0.45\textwidth]{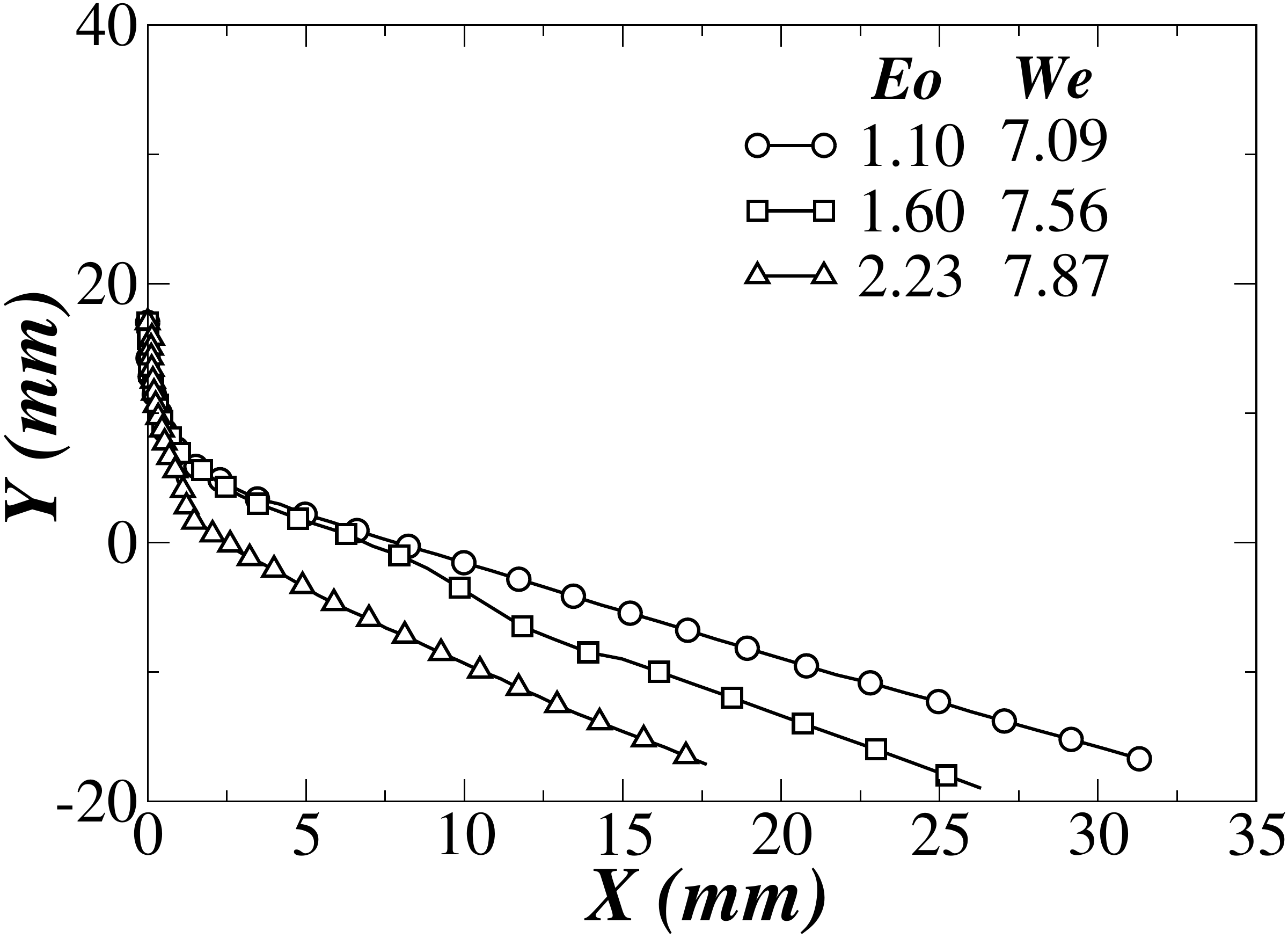} \hspace{2mm} \includegraphics[width=0.45\textwidth]{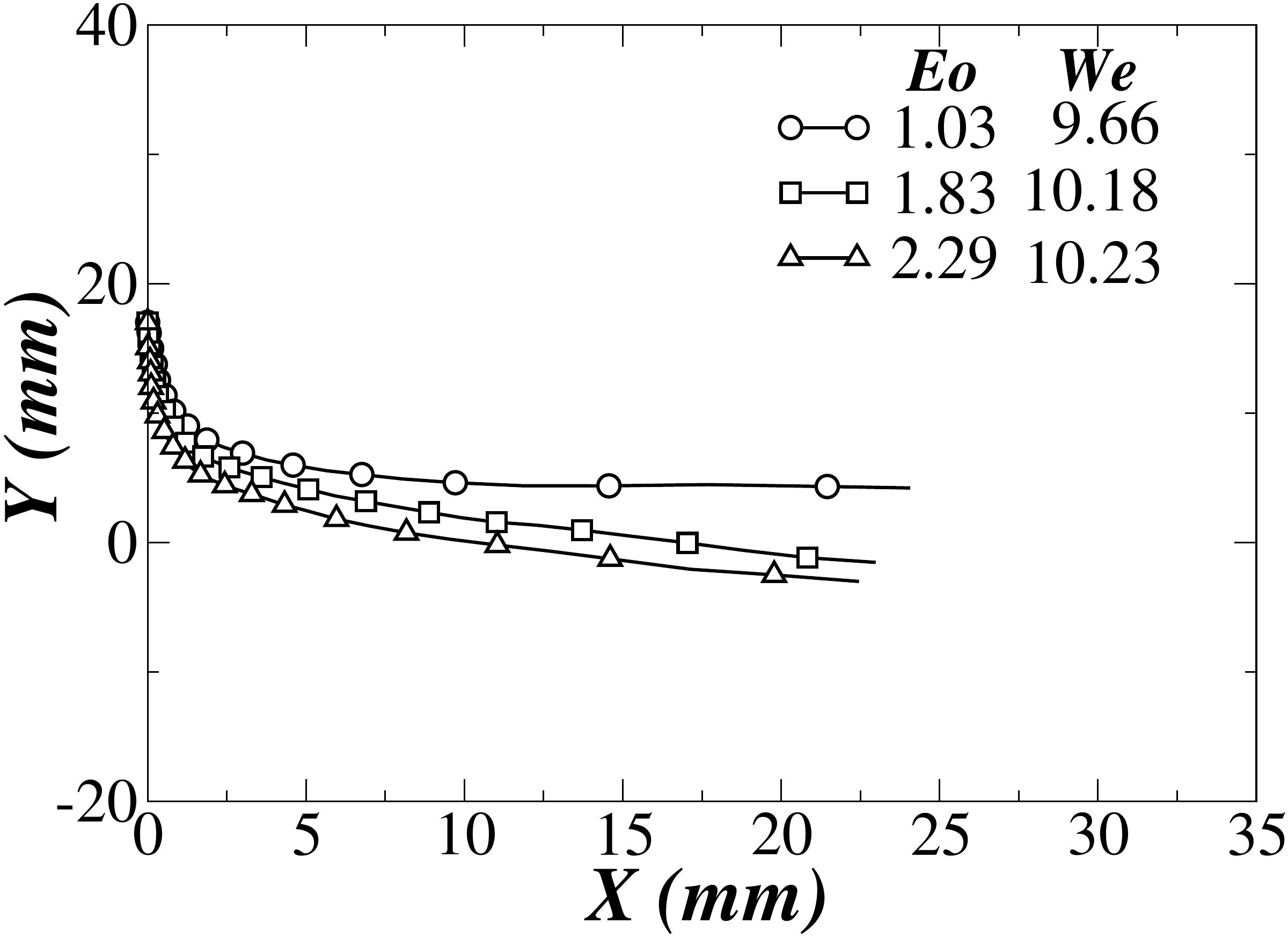} \\
        \hspace{0.5cm} {\large (c)} \hspace{7.8cm} {\large (d)}\\
    \includegraphics[width=0.45\textwidth]{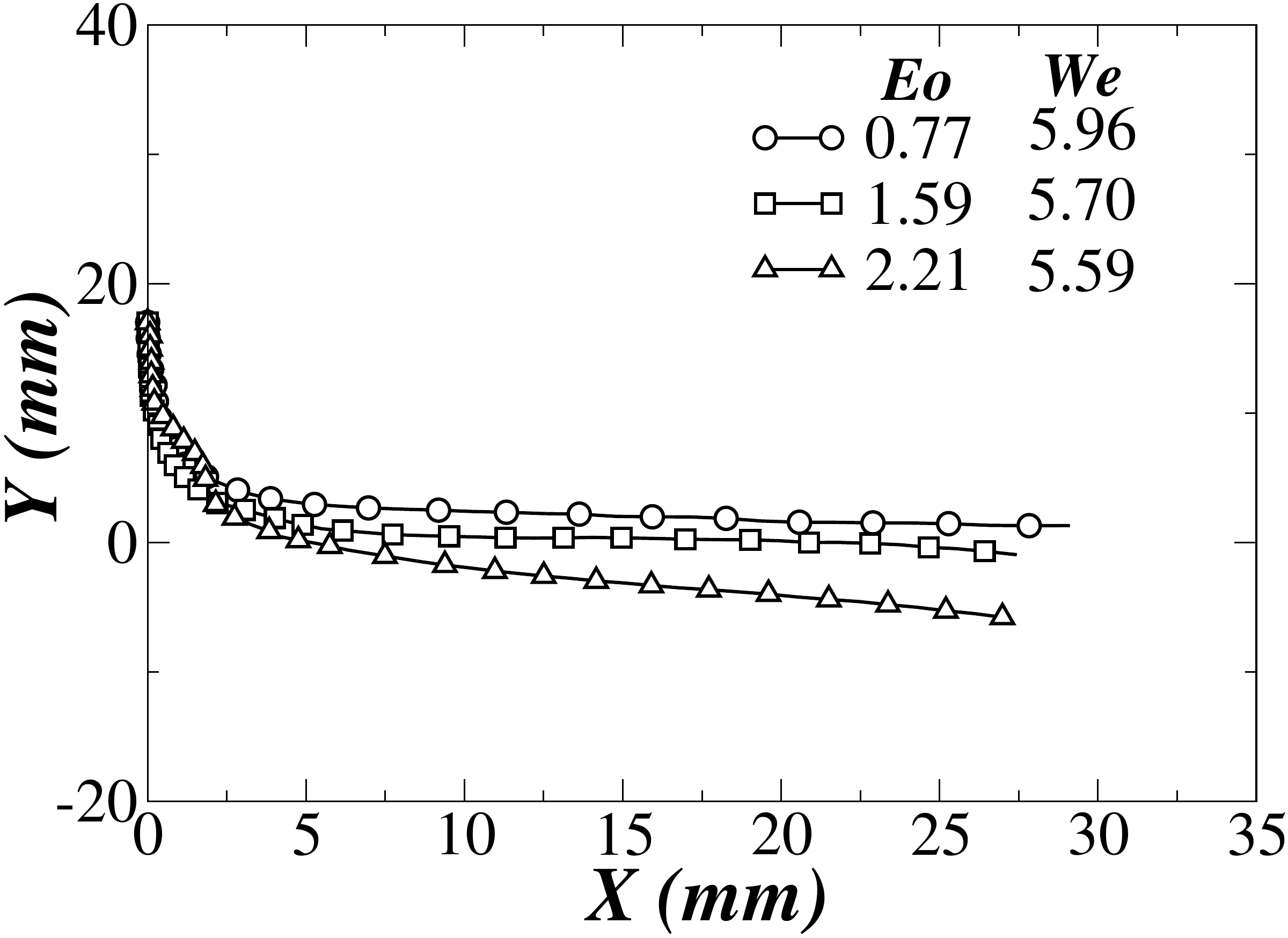} \hspace{2mm} \includegraphics[width=0.45\textwidth]{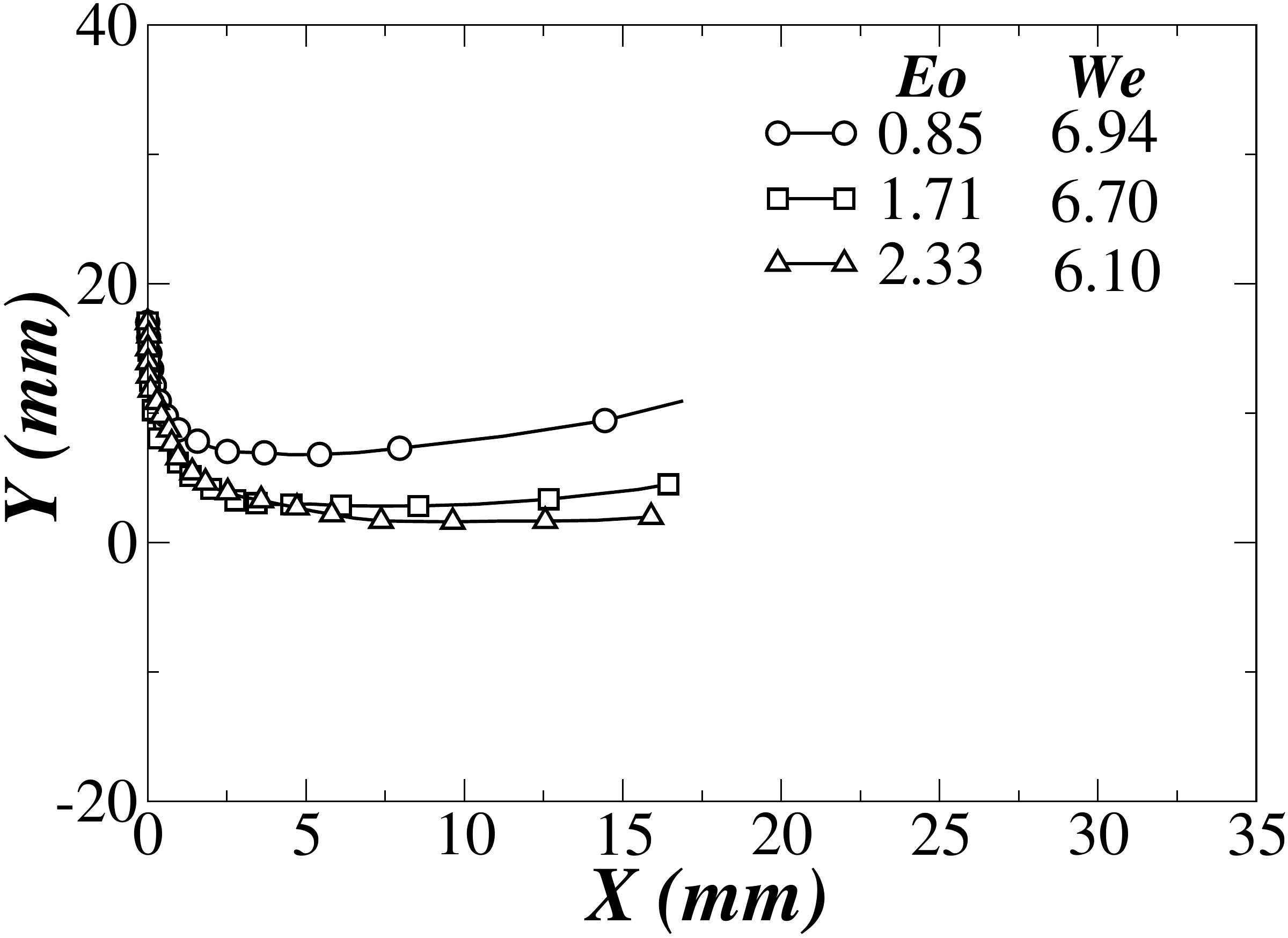} \\
        \hspace{0.5cm} {\large (e)} \hspace{7.8cm} {\large (f)}\\
     \includegraphics[width=0.45\textwidth]{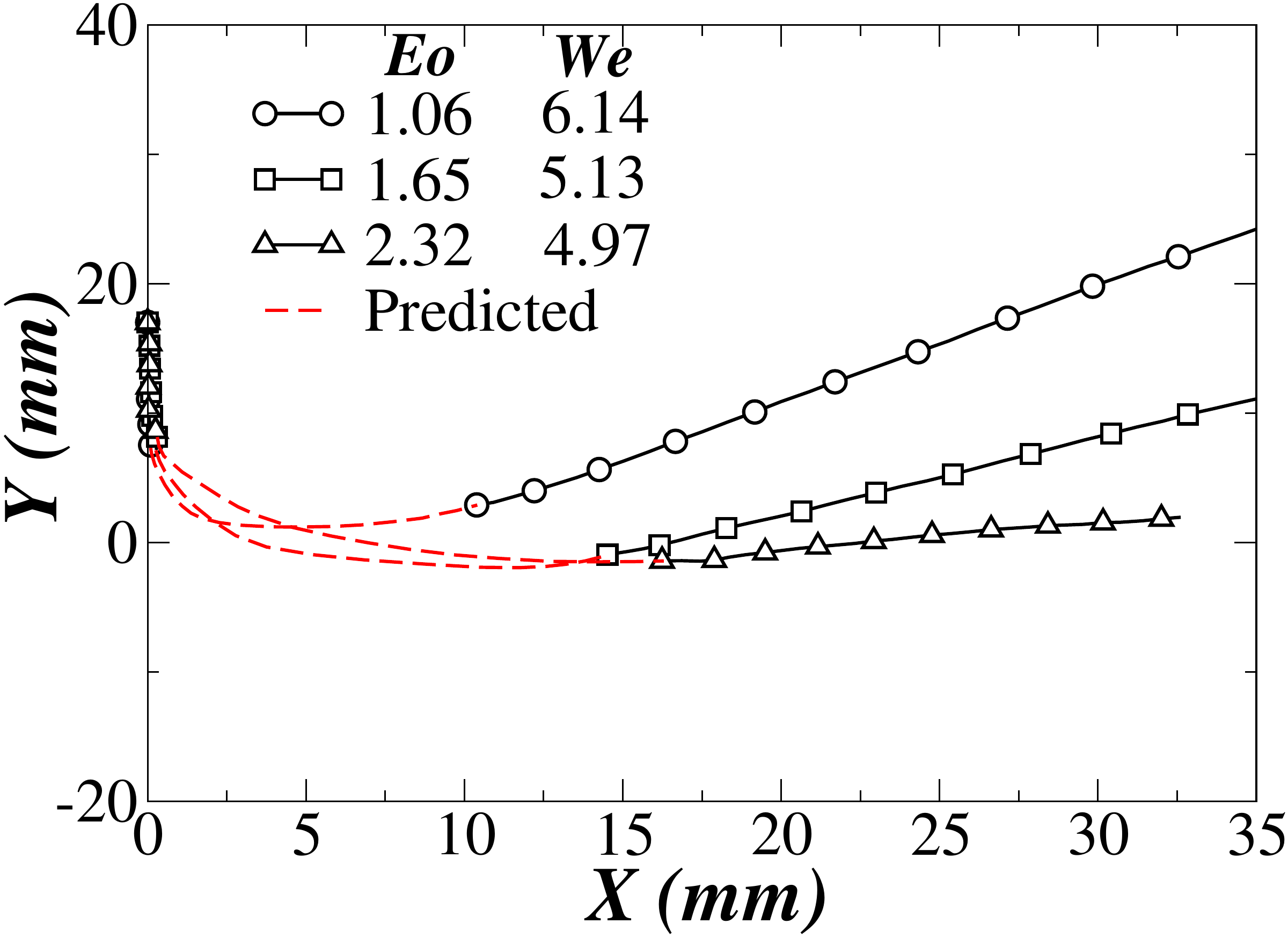} \hspace{2mm} \includegraphics[width=0.45\textwidth]{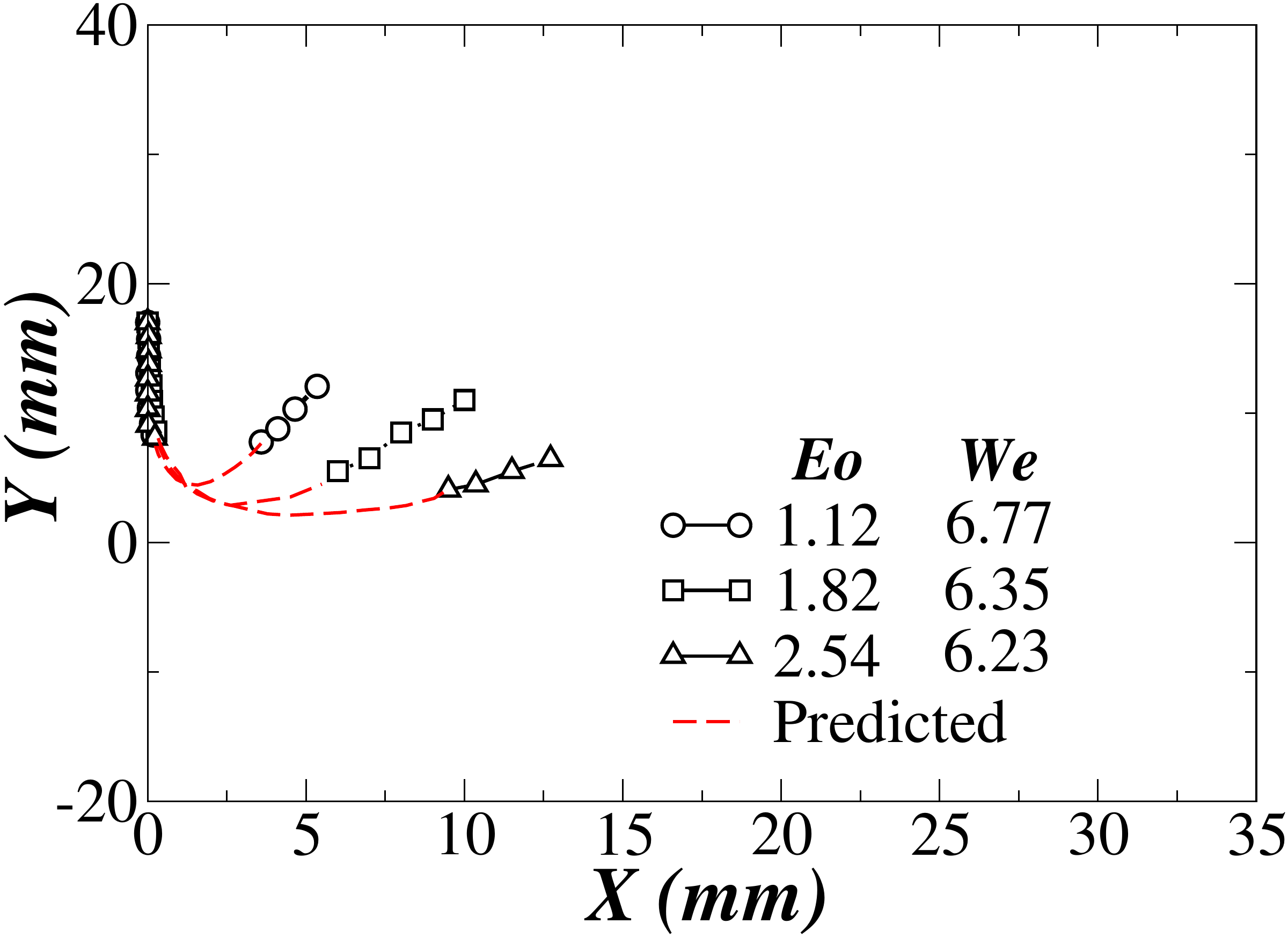} \\
    \caption{Trajectories of a water droplet ($Oh \approx 0.0015$) exhibiting (a,c,e) the vibrational and (b,d,f) the bag breakup modes for (a,b) $\alpha=0^\circ$, (c,d) $\alpha=30^\circ$ and (e,f) $\alpha=60^\circ$.  }
    \label{fig8}
\end{figure}

Next we investigate the path followed by the droplet subjected to an oblique air flow for different sets of $Eo$ and $We$ as it plays important role in combustors designing and optimisation. Figs. \ref{fig8}(a,b), (c,d) and (e,f) correspond to $\alpha=0^\circ$ (cross-flow configuration), $\alpha=30^\circ$ and $\alpha=60^\circ$, respectively. Figs. \ref{fig8} (a,c,e) and (b,d,f) show the trajectories for the droplet exhibiting the vibrational mode and the bag breakup, respectively. Here, $Y=0$ represents the centre line of the nozzle. The trajectories are plotted before fragmentation of the droplet. It can be seen that droplet travel a short distance from the nozzle and undergoes a bigger turning in the bag breakup cases as compared to the vibrational cases. It is observed that for all $\alpha$ values considered, when the droplet velocity vector (tangent to the trajectory) aligns closer to the air flow direction, the drag force exerted on the droplet is more, which increases the stretching of the droplet, and the surface tension force can no longer keep the droplet intact and it disintegrates into smaller droplets. Increasing the droplet size (increasing $Eo$) increases the curvature of the droplet trajectory. It is to be noted here that the value of $We$ considered is close to the critical Weber number for each case. {However, the fact that the larger droplets, although smaller than the critical diameter based on the capillary length scale, break up even when the trajectory of the droplet is not aligned with the air stream direction. This indicates that the gravity force, aiding into droplet distortion, leads to increase the drag to overcome the cohesive force that keeps the droplet intact.}

 \begin{figure}[H]
    \centering
    \hspace{1cm} {\large (a) $\alpha=0^\circ$} \hspace{3cm} {\large (b) $\alpha=30^\circ$} \hspace{3cm} {\large (c) $\alpha=60^\circ$}\\
    \includegraphics[width=0.3\textwidth]{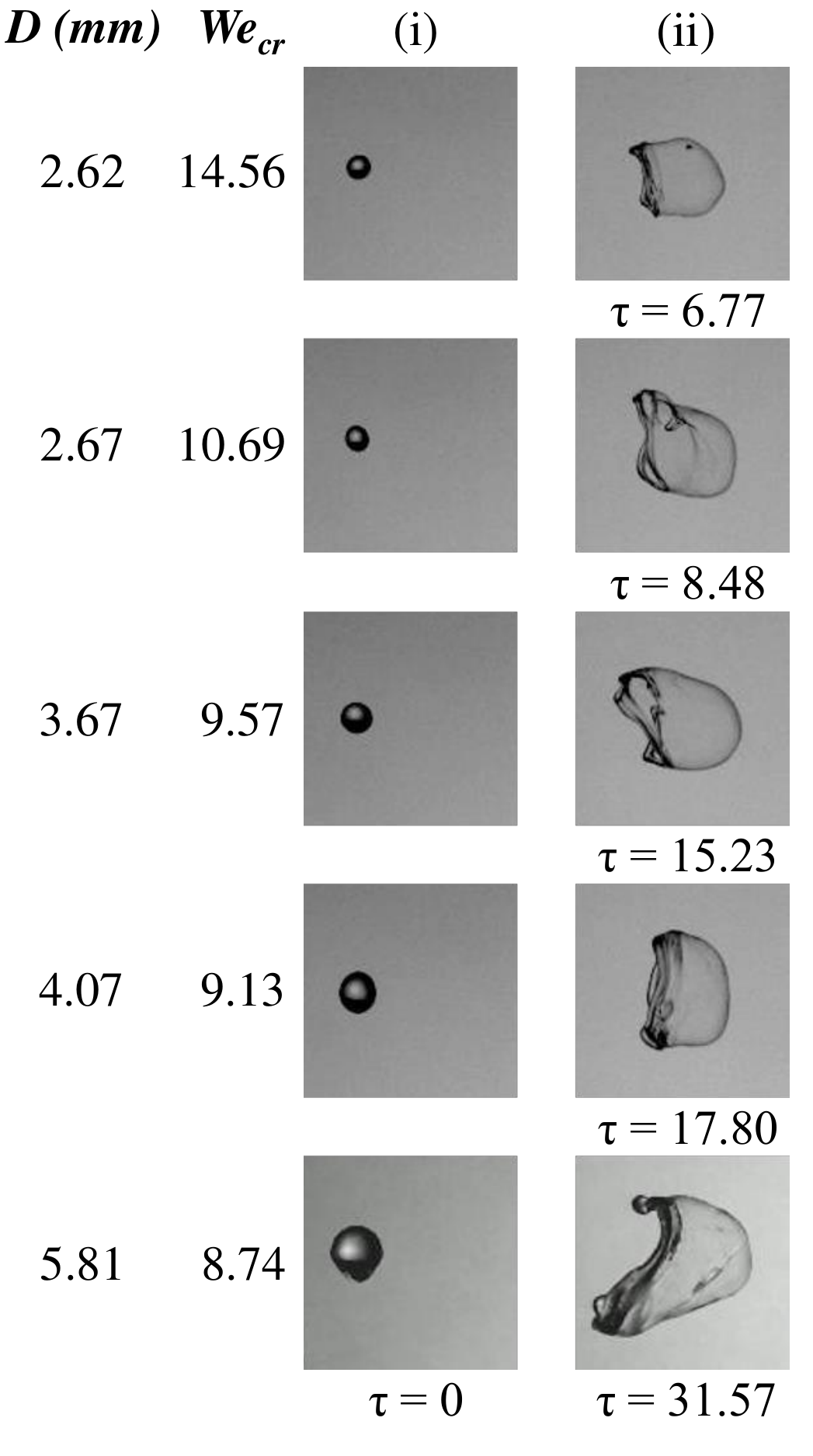}
    \includegraphics[width=0.31\textwidth]{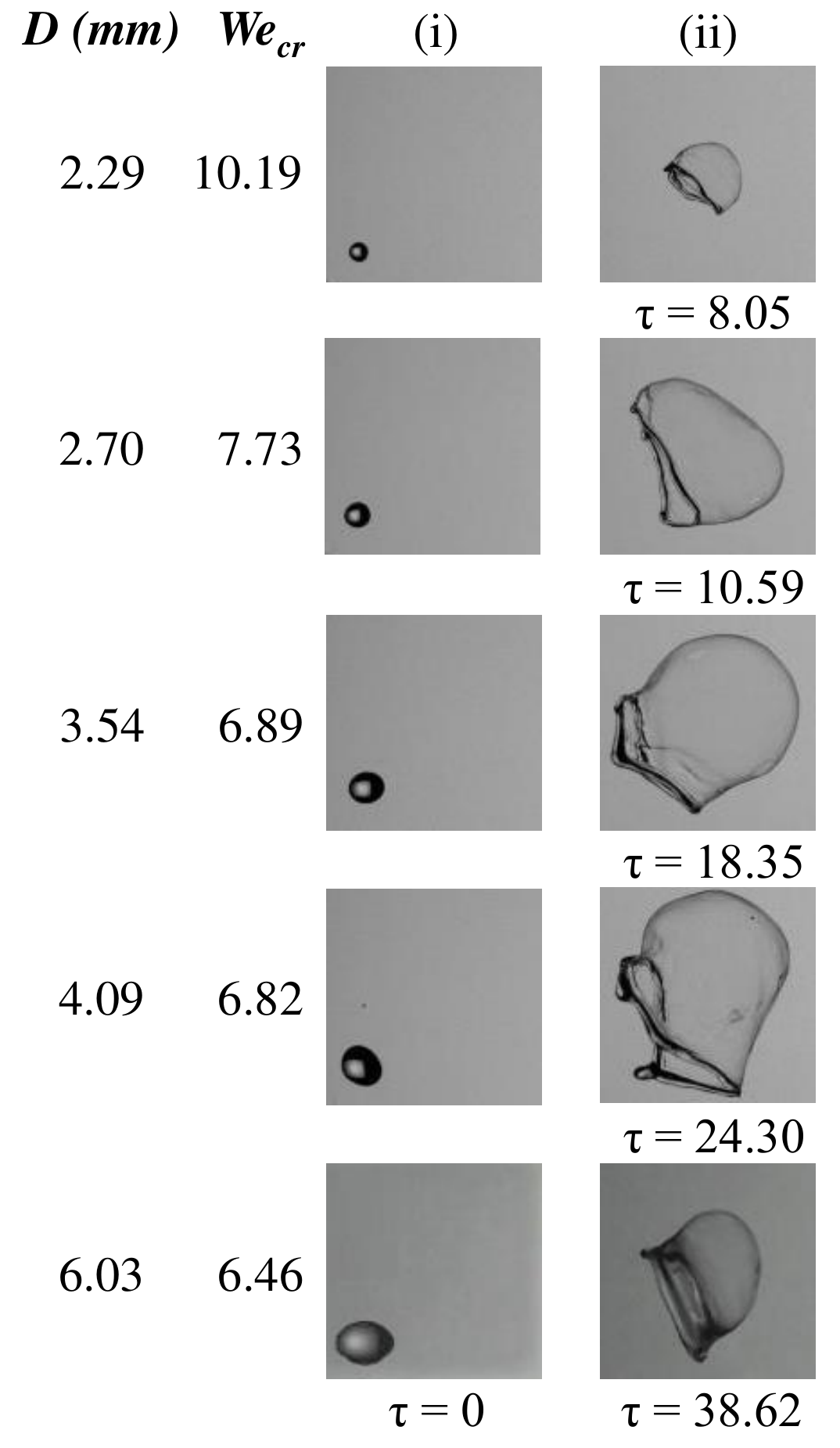}
    \includegraphics[width=0.31\textwidth]{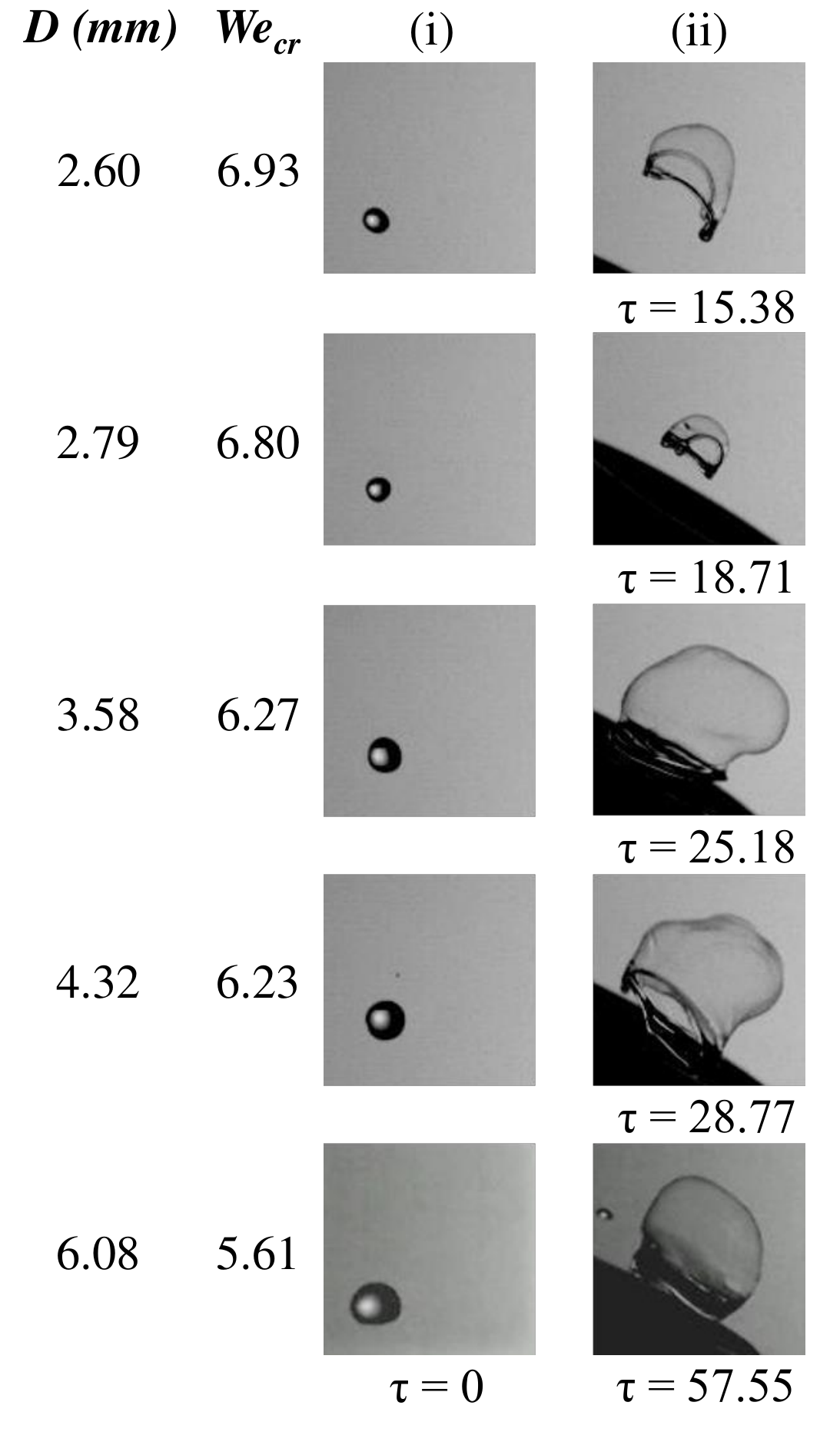}
    \caption{Breakup of a water droplet ($Oh\approx0.0015$) for (a) $\alpha=0^\circ$, (b)  $\alpha=30^\circ$ and (c)  $\alpha=60^\circ$. For each angle, the left side images are corresponding to drop shape just before reaching the air stream, and the right images are the same droplet corresponding to $\tau$ (written below the images) just before the onset of the bag breakup.}
    \label{fig11a}
\end{figure}

    Figs. \ref{fig11a}(a), (b) and (c) show instantaneous snapshots of the droplet of different sizes just before entering the air stream (panel (i); $\tau=0$) and just before the onset of bag breakup (panel (ii); the value of $\tau$ is written below the droplets) for $\alpha=0^{\circ}$, $30^{\circ}$ and $60^{\circ}$, respectively. The value of $\tau$ mentioned in panel (ii) for each $\alpha$ essentially represents the time required to complete the bag formation for different sizes of the droplet at its critical Weber number, $We_{cr}$. It can be observed that for each value of $\alpha$, increasing the droplet size increases the value of the dimensionless breakup time. Increasing $Eo$ (increasing droplet size) decreases the value of $We_{cr}$, and increases the breakup time because of the decrease in the critical air velocity requirement for the larger droplets. It can also be seen that with the increase in the obliquity angle of the air nozzle, the dimensionless breakup time, $\tau$ shows a significant increase. This can be attributed to the droplet deceleration prior to aligning closely with the air jet direction. Interestingly, the droplet breakup occurs much closer to the air nozzle for high obliquity angle ($\alpha$ = 60$^\circ$) with significantly larger residence time. This is an important requirement from the design perspective of a swirl stabilised combustors as the flame would anchor much closer to the swirler warranting accelerating swirl flow (or converging) to increase flame standoff distance. Such a practice have been adopted in previous studies (see Refs. \cite{merkle2003effect,gupta2001swirl}). For $\alpha$ = 0$^\circ$ (cross-flow configuration), the path traveled by the droplet is large as compared to the path traveled by the droplet at any other obliquity angle conditions. Note that the droplet traveled more distance from the nozzle in $\alpha=0^\circ$ case than other cases with $\alpha$ values, but still the non-dimensional time is lower, which can be attributed to the acceleration of the droplet in the cross-flow condition. This is unlike the oblique cases where the deceleration of droplet is not observed as droplet motion is not totally against gravity. 
    
\begin{figure}[H]
    \centering
    \includegraphics[width=0.25\textwidth]{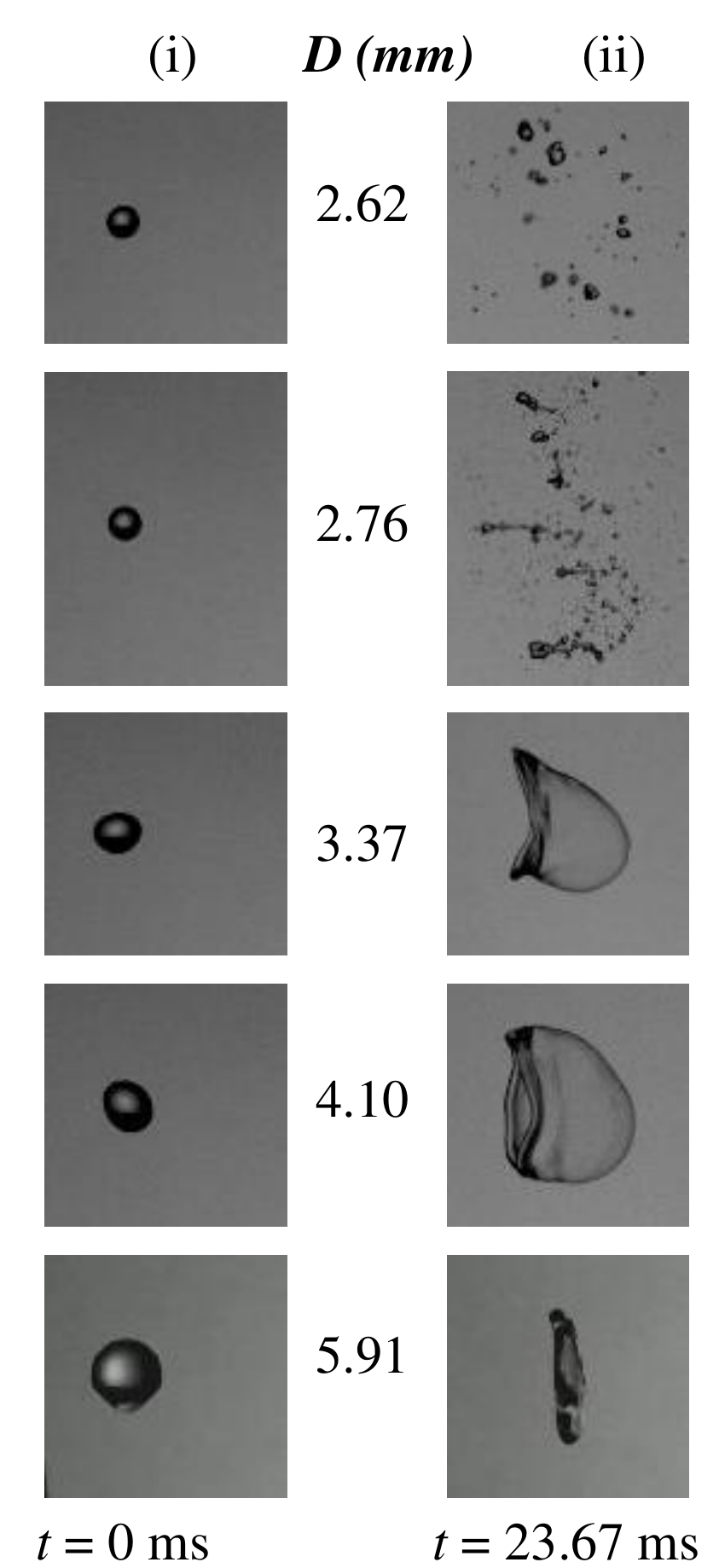}
    \caption{Breakup of a water droplet ($Oh\approx0.0015$) at $\alpha=0$ (cross-flow configuration). The left side images are corresponding to drop shape just before entering the air stream and the right set of images are for the same droplet $t=23.67$ ms.}
    \label{fig11b}
      \end{figure}

    In order to demonstrate that small droplets require less time to disintegrate as compared to bigger droplets at their respective values of the critical Weber number, in Fig. \ref{fig11b}, we present the shapes of different size droplets at two dimensional times, $t=0$ and $t=23.67$ ms in the cross-flow configuration ($\alpha=0$). It can be seen that the droplets are almost spherical when they enter the air stream. We observed that the small droplets (with $D=2.62$ mm and 2.76 mm) at their critical Weber numbers completely disintegrate, the intermediate size droplets ($D=3.37$ mm and 4.10 mm) form bags and the bigger droplet ($D=5.91$ mm) appears as a disk at $t=23.67$ ms. Although the bigger droplet breaks at smaller Weber number, the time required for the breakup increases with the droplet size. As in this case, the velocity requirement is lower, the rate of energy transfer to the droplet from the air stream is reduced resulting into larger residence time to overcome the effect due to the surface tension and the viscous forces.

\begin{figure}[H]
    \centering
    \hspace{1cm} {\large (a) $\alpha = 0^\circ$} \hspace{6cm}  {\large (b) $\alpha = 10^\circ$}
    \vspace{5mm}
    \includegraphics[width=0.45\textwidth]{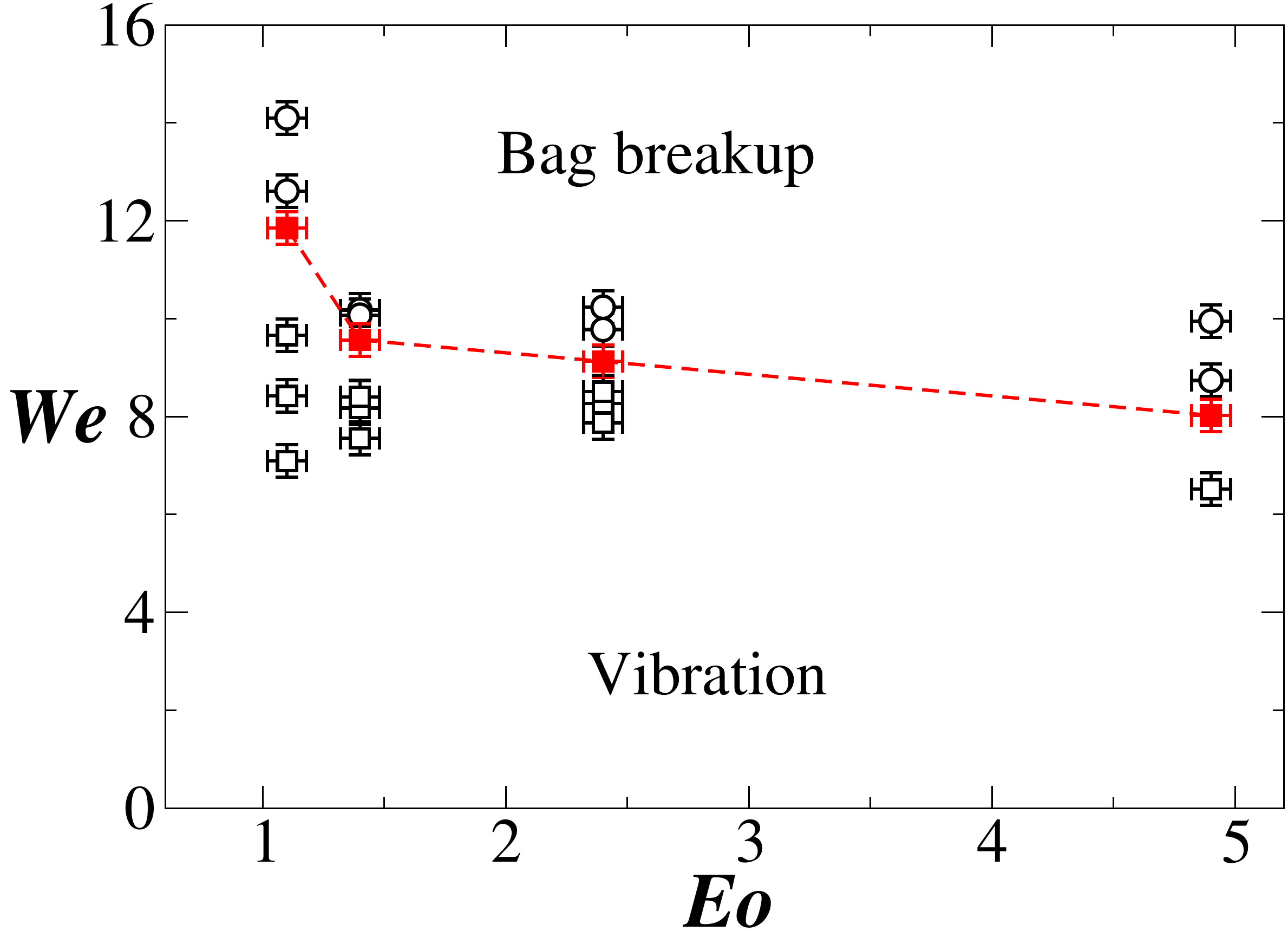} \hspace{2mm} \includegraphics[width=0.45\textwidth]{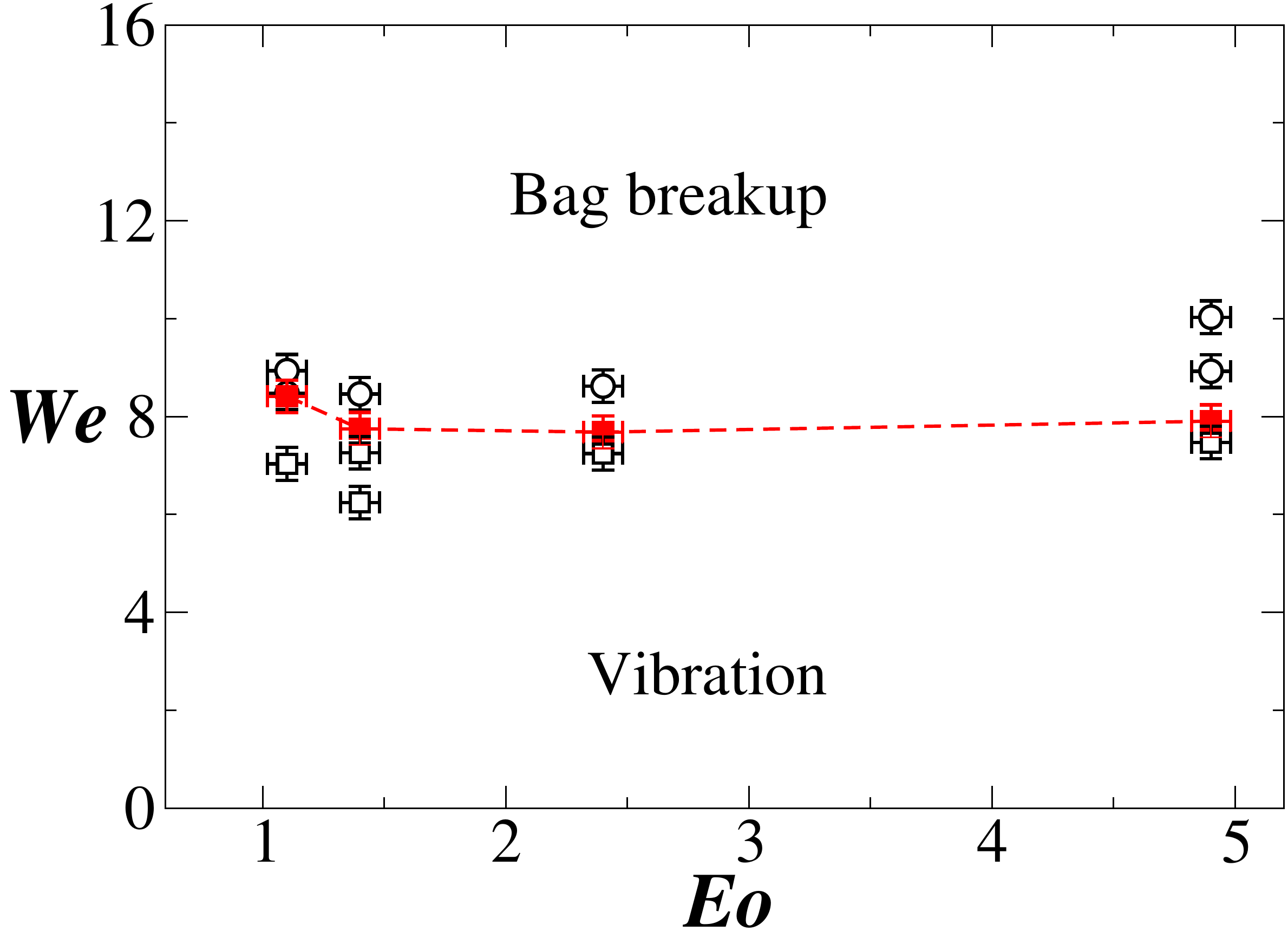} \\
    \vspace{1mm}
    \hspace{1.cm}  {\large (c) $\alpha = 20^\circ$} \hspace{6cm}  {\large (d) $\alpha = 30^\circ$} \\
    \includegraphics[width=0.45\textwidth]{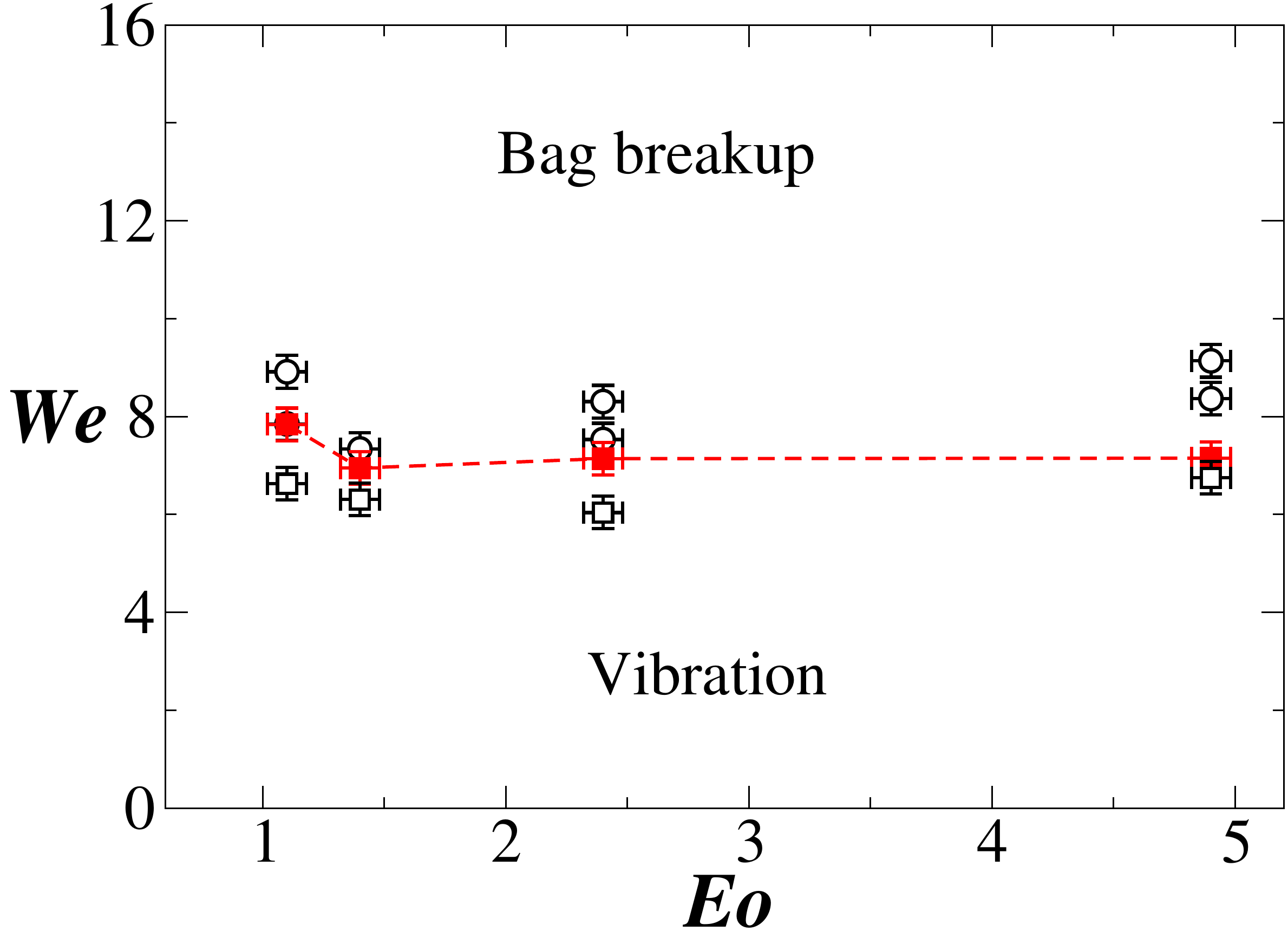} \hspace{2mm} \includegraphics[width=0.45\textwidth]{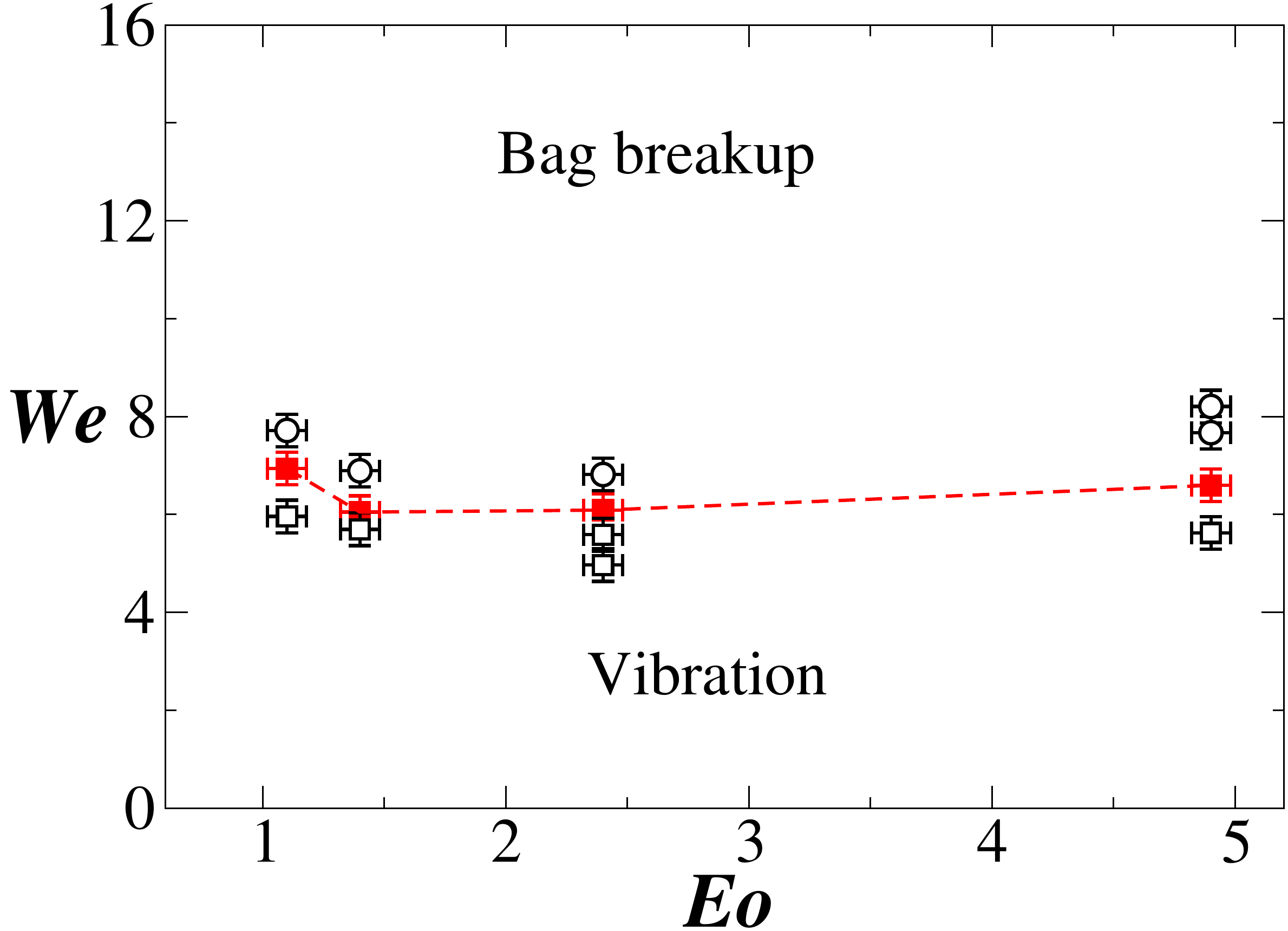} \\
    \vspace{3mm}
    \hspace{1cm}  {\large (e) $\alpha = 40^\circ$} \hspace{6cm}  {\large (f) $\alpha = 60^\circ$} 
    \vspace{4mm}
    \includegraphics[width=0.45\textwidth]{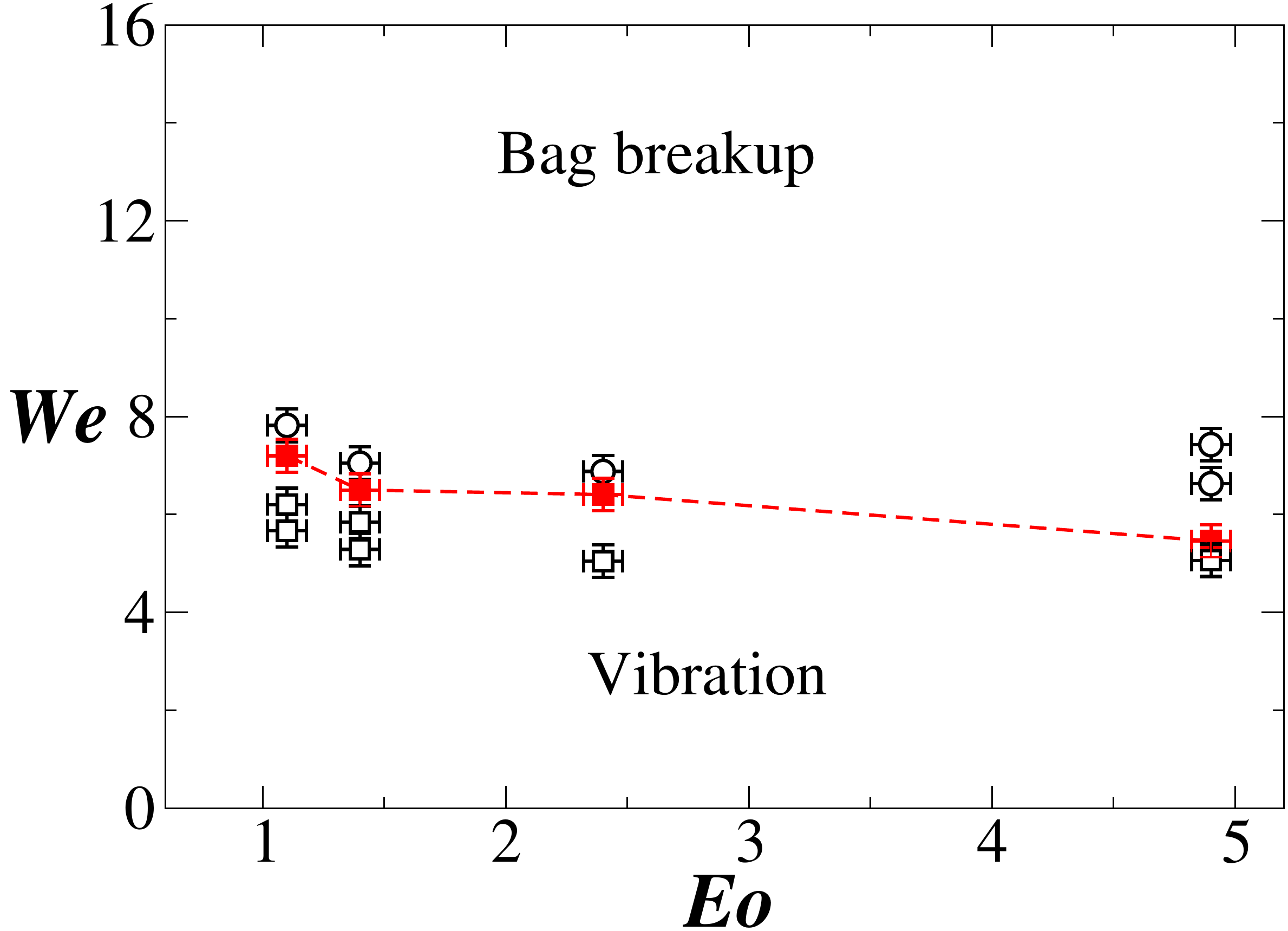} \hspace{2mm} \includegraphics[width=0.45\textwidth]{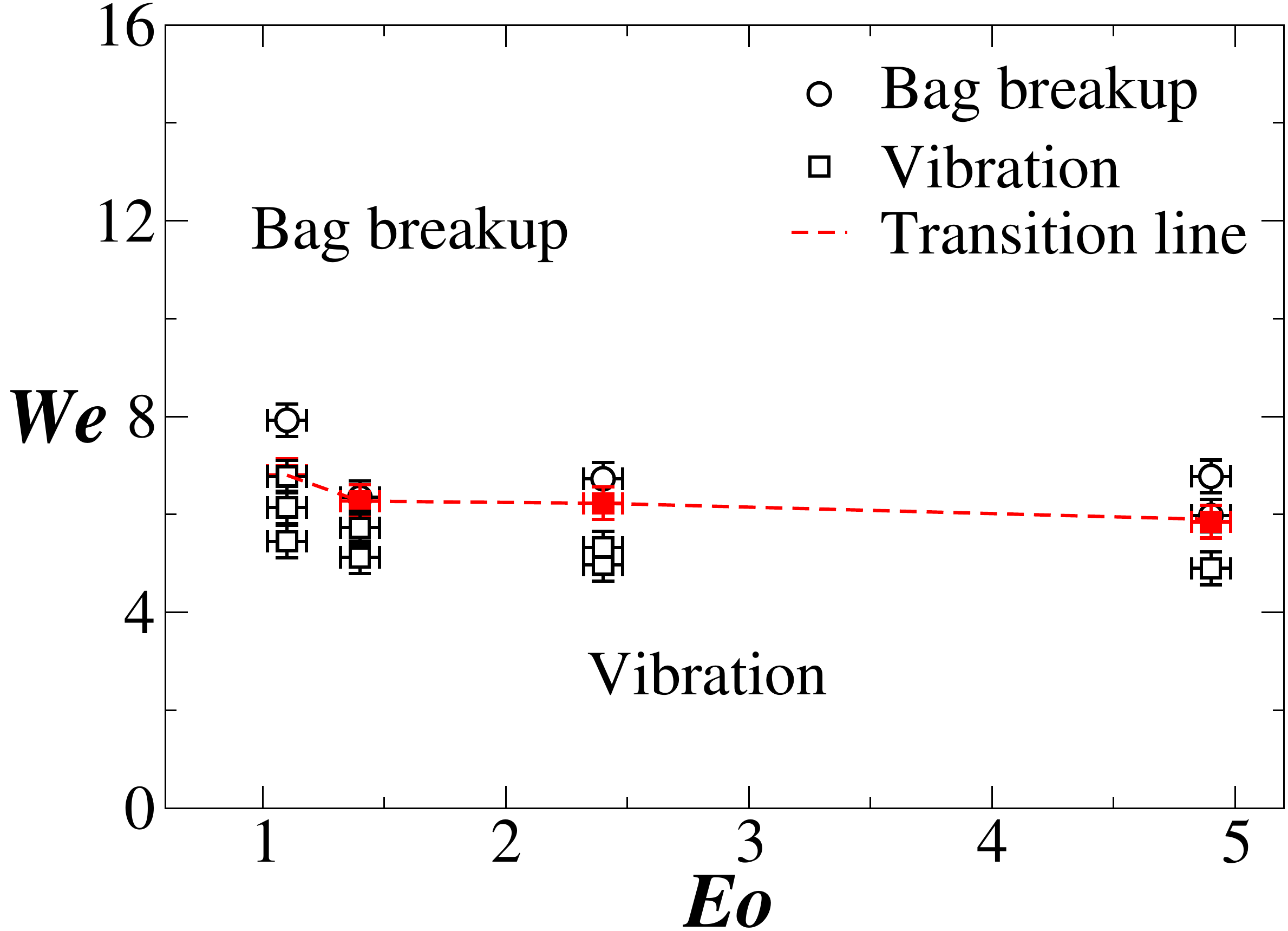} \\
    \caption{Region maps showing the vibrational and the bag breakup modes in $We - Eo$ plane for a water droplet ($Oh\approx0.0015$) at different angles of inclination of the air stream, $\alpha$. The red dashed line in each panel shows the transition between the vibrational and bag breakups.}
    \label{fig6}
\end{figure}

Then we investigate the effect of angle of the air stream, $\alpha$ on the critical Weber number $We_{cr}$. In Fig. \ref{fig6}, we present the region maps showing the vibrational and the bag breakup modes in the $We-Eo$ plane for different values of $\alpha$. Here, the $Eo$ is changed by changing the droplet size in our experiments. In each panel, the red dashed line with filled square symbols depicts the boundary between the vibrational (open circle symbols) and the bag breakup (open square symbols) regimes. For $\alpha=0^\circ$ (cross-flow configuration), at $Eo=1.1$, the value of the $We_{cr} \approx 14$, which conforms the finding of previous studies \cite{wierzba1990deformation,hanson1963}. Increasing the value of $Eo$ decreases the value of $We_{cr}$ initially and then decreases at a slower rate. For high values of $Eo$, $D \ge l_c$. Thus the gravity force starts to dominate the surface tension force, and the droplet deforms from its initially spherical shape. This, in turn, increases the drag force exerted on droplet due to the increase in its surface area, which helps the droplet to deform more  \cite{jalaal2012fragmentation}. Thus, the critical Weber number decreases with the increase in $Eo$ (or size of the droplet). Note that in the cross-flow condition ($\alpha=0^\circ$), air flows in the horizontal direction and the component of net acceleration in the vertical direction remains as $g$. Hence the reported E\"{o}tv\"{o}s number is calculated based on the gravitational acceleration component only.

\begin{figure}[H]
        \centering
        \includegraphics[width=0.5\textwidth]{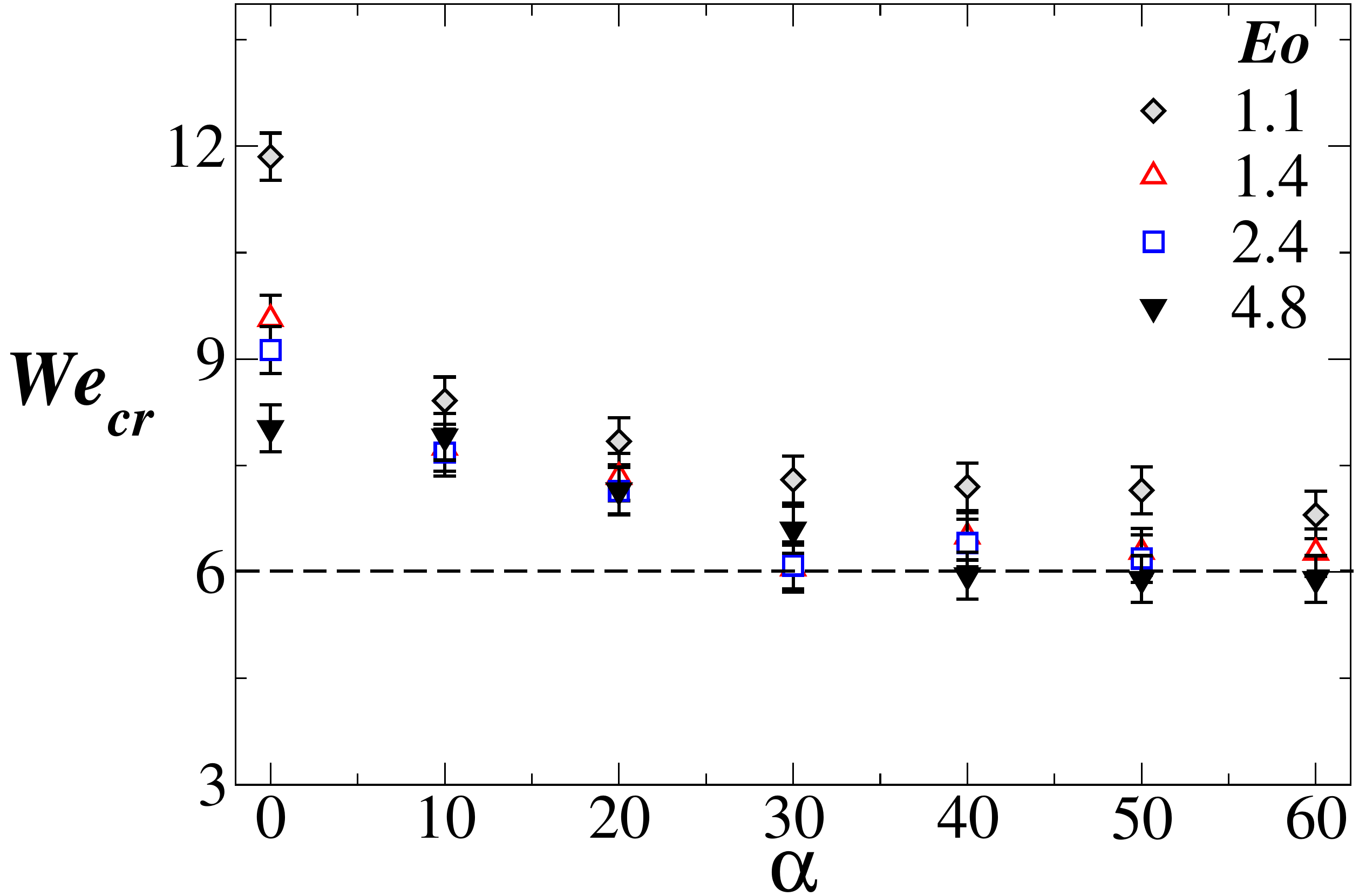}
        \caption{Variation of the transitional Weber number, $We_{cr}$ with $\alpha$ (in degree) for water droplet ($Oh\approx0.0015$) and  different values of E\"{o}tv\"{o}s number. The uncertainty in the measurement in angles is $\pm0.5^\circ$. }
        \label{fig7}
    \end{figure}

For the cases with the oblique air flow, the droplet undergoes a curvilinear motion with a rapid deceleration followed by acceleration with a small radius of curvature in its trajectory (see Fig. \ref{fig:1a}). The curvilinear trajectory makes the internal pressure distribution to fluctuate as the droplet is continuously changing its direction. The pressure variation in a curvilinear motion is normal to the instantaneous velocity vector, which increases the stretching of the droplet, and hence the frontal area and the drag force also increase. Comparison of the results presented in Fig. \ref{fig6}(a)-(f) reveals that increasing the value of $\alpha$ decreases the value of $We_{cr}$. For all values of $\alpha$ considered, $We_{cr}$ drops sharply as we increase $Eo$ initially and then decreases slightly or remains constant. This behaviour can also be seen in Fig. \ref{fig7}, where the variations of $We_{cr}$ versus $\alpha$ are plotted for different values of $Eo$. It can be seen that $We_{cr}$ decreases to an asymptotic value of 6 (approximately). Interestingly, Villermaux and Bossa \cite {Villermaux2009} also theoretically found that $We_{cr} \approx 6$ for $\alpha = 90^\circ$ (in case of the opposed flow configuration). It can be seen that Fig. \ref{fig:1a} that at $\alpha$ = 60$^\circ$ the droplet almost aligned with the direction of air stream. This is the reason why $We_{cr}$ for $\alpha$ = 60$^\circ$ is approximately equal to 6, like in the case of the opposed flow configuration considered by Villermaux and Bossa \cite {Villermaux2009}.

\begin{figure}[H]
\centering
\hspace{0.5cm} {\large (a)} \hspace{7.8cm} {\large (b)}\\
\includegraphics[width=0.45\textwidth]{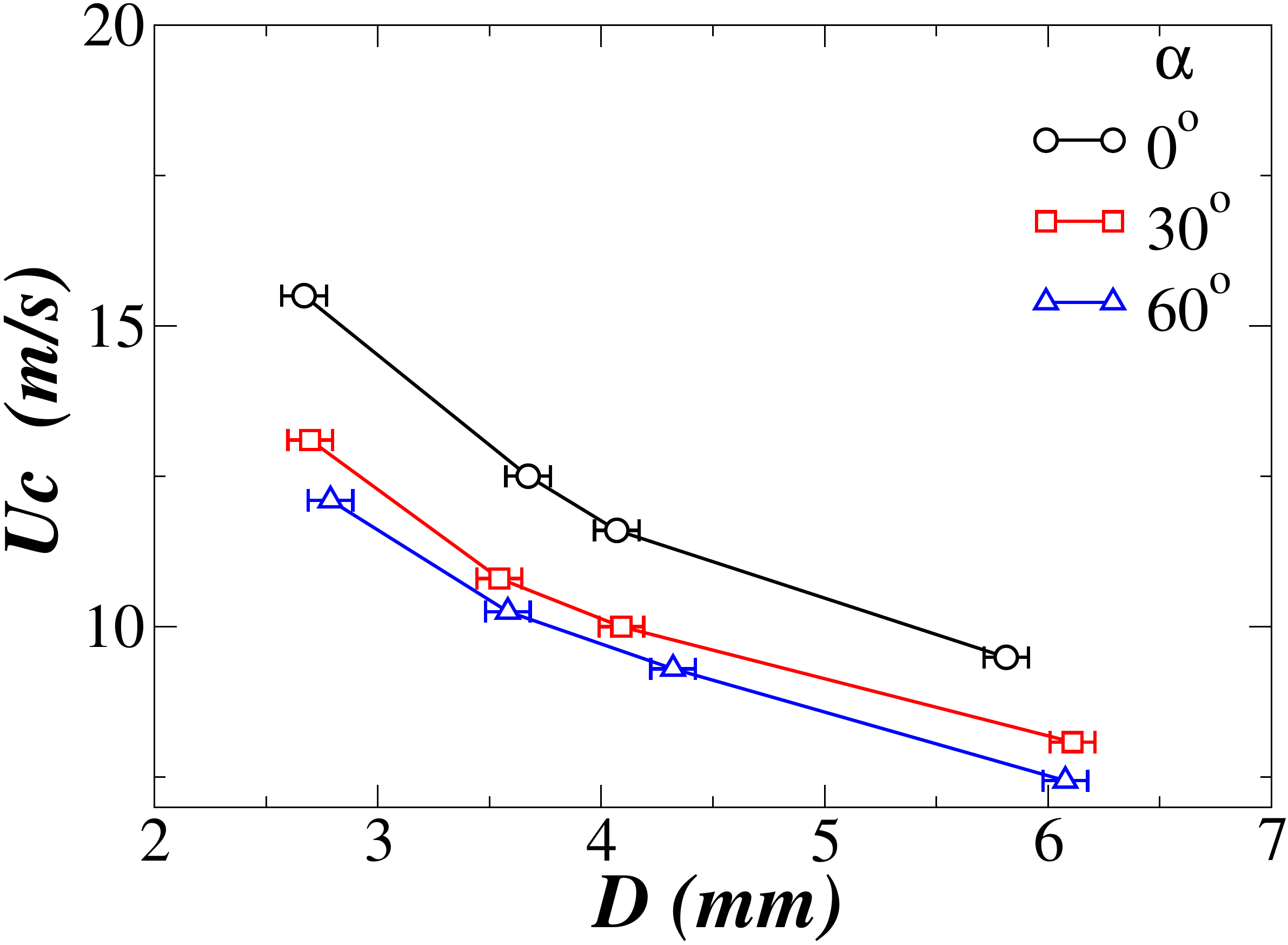} \hspace{0mm} \includegraphics[width=0.46\textwidth]{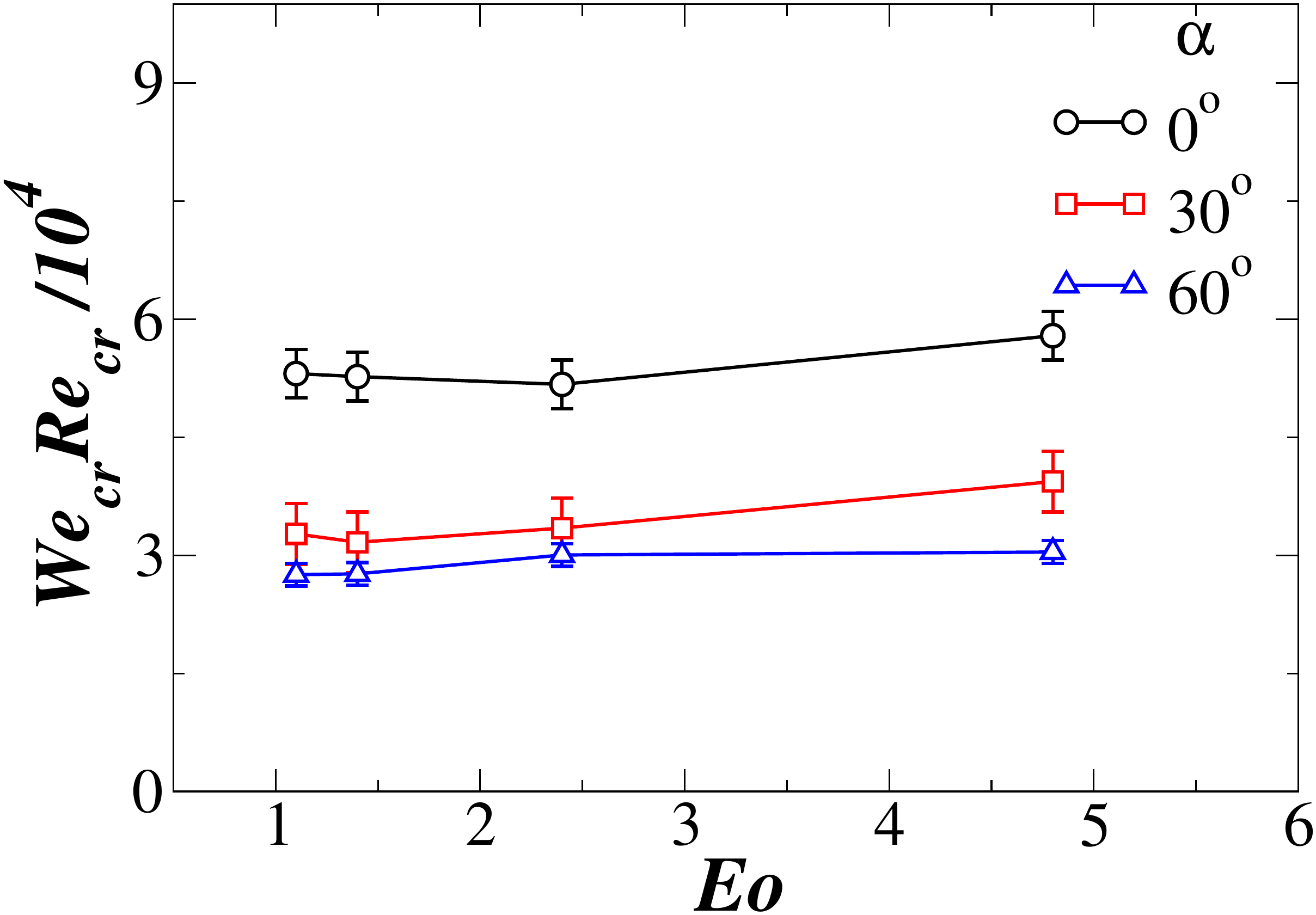} \\
\hspace{0.5cm} {\large (c)} \hspace{7.8cm} {\large (d)}\\
\includegraphics[width=0.45\textwidth]{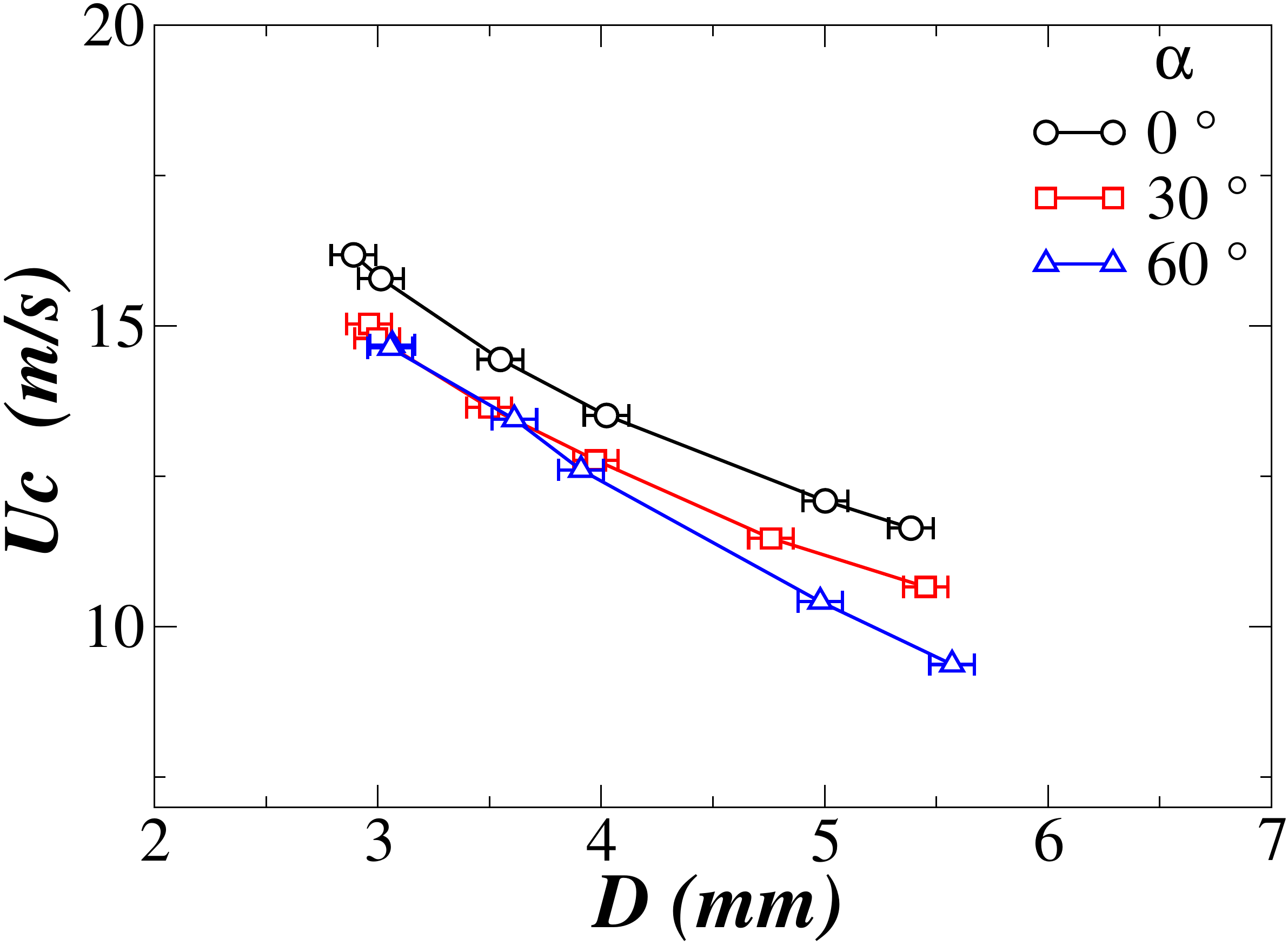} \hspace{2mm} \includegraphics[width=0.47\textwidth]{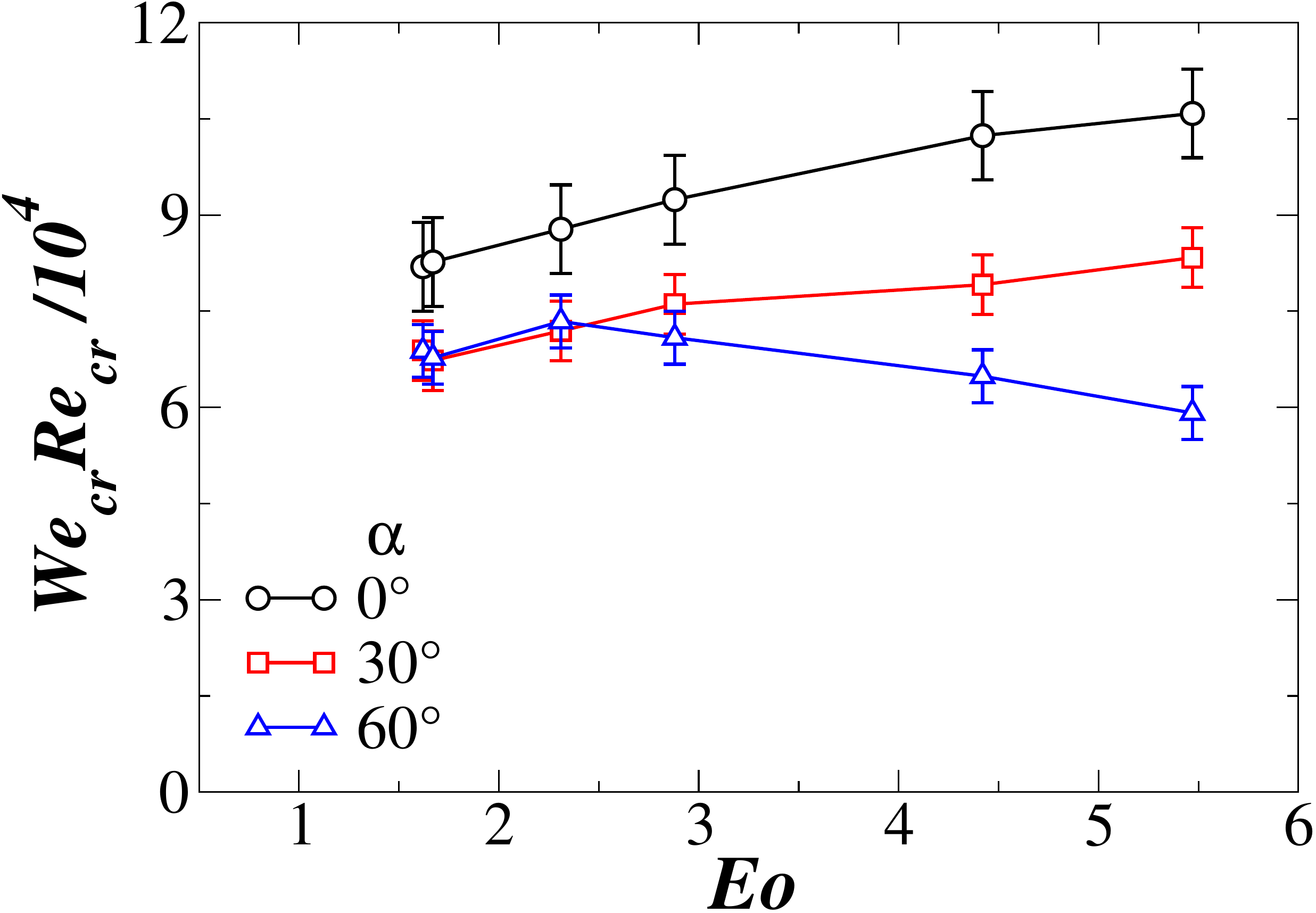} \\
\caption{ (a,c) The critical dimensional air velocity, $U_c$ (m/s) versus $D$ (mm), and (b,d) $We_{cr}Re_{cr}$ versus $Eo$ for different values of $\alpha$. Panels (a,b) and (c,d) correspond to the droplets of water ($Oh\approx0.0015$) and 80\% glycerol + 20\% water solution ($Oh\approx0.078$), respectively.}
\label{fig4}
\end{figure}

In Figs. \ref{fig4}(a) and (c), the critical air stream velocity, $U_c$ that is just sufficient for the bag breakup is plotted against the droplet diameter, $D$ for three different values of $\alpha$ with droplets of water and 80\% glycerol+ 20 \%water solution, respectively. It can be seen that increasing the size of the droplet decreases the value of the velocity of the air stream required for the bag breakup. This behaviour can be attributed to the large frontal cross-section area and the large drag coefficient due to the deformation associated with bigger droplets. Additionally, the wake region \cite{flock2012} developed behind the deformed droplet facilitates the skewed internal pressure distribution leading to excessive stretching. A similar finding was also reported in Refs. \cite{wierzba1990deformation,chou1998temporal}. {Close inspection of Fig. \ref{fig4}(a) and (c) also reveals that increasing the viscosity of the droplet the critical velocity requirement increases slightly for the same size of the droplet, which is also observed by previous researchers \cite{pilch1987,guildenbecher2009,hanson1963}. This is due to the suppression of the surface instability as the viscosity is increased. It can be seen that increasing $\alpha$ for the same droplet size decreases the critical velocity requirements for the droplet to undergo the bag breakup.} This can be attributed to the effective energy dissipation to overcome the surface tension and the viscous effects through the larger turning in the droplet trajectory. Note that the rate of change of the angular momentum provides a measure of the energy transfer from air stream to the droplet to overcome the viscous and surface energy dissipation. Thus decreasing the radius of curvature in the curvilinear trajectory of the droplet increases the change in its angular momentum leading to a more effective energy transfer in accordance with the conservation of energy principle.

It is obvious that as the droplet size increases, the critical Weber number ($We_{cr}$) decreases but the corresponding critical Reynolds number ($Re_{cr}$) of the air stream increases even though the air velocity is decreased (except when the stretching is excess). However, in Fig. \ref{fig4}(b) and (d), it can be seen that the product of $We_{cr}$ and $Re_{cr}$ does not vary much with the increase in the droplet size. This is observed for all angles of the air stream. The product of $We_{cr}$ and $Re_{cr}$ signifies the inertial force requirement to overcome the viscous and the surface tension forces, such that a low value of $We_{cr} \times Re_{cr}$ represents a situation when a less inertia force can lead to the droplet breakup. Increasing $\alpha$ decreases $We_{cr} \times Re_{cr}$ as the curvature in the droplet trajectory increases with increasing $\alpha$ leading to more effective energy transfer from the air stream to the droplet to overcome viscous and surface tension forces. Hence, it can be concluded that the opposed flow configuration is an effective way to promote the secondary atomization at a low cost of input energy \cite{merkle2003effect,gupta2001swirl}).

\begin{figure}[H]
\centering
\hspace{1cm} {\large (a) $\alpha = 0^\circ$} \hspace{6.5cm}{ \large (b) $\alpha = 30^\circ$}\\
\includegraphics[width=0.45\textwidth]{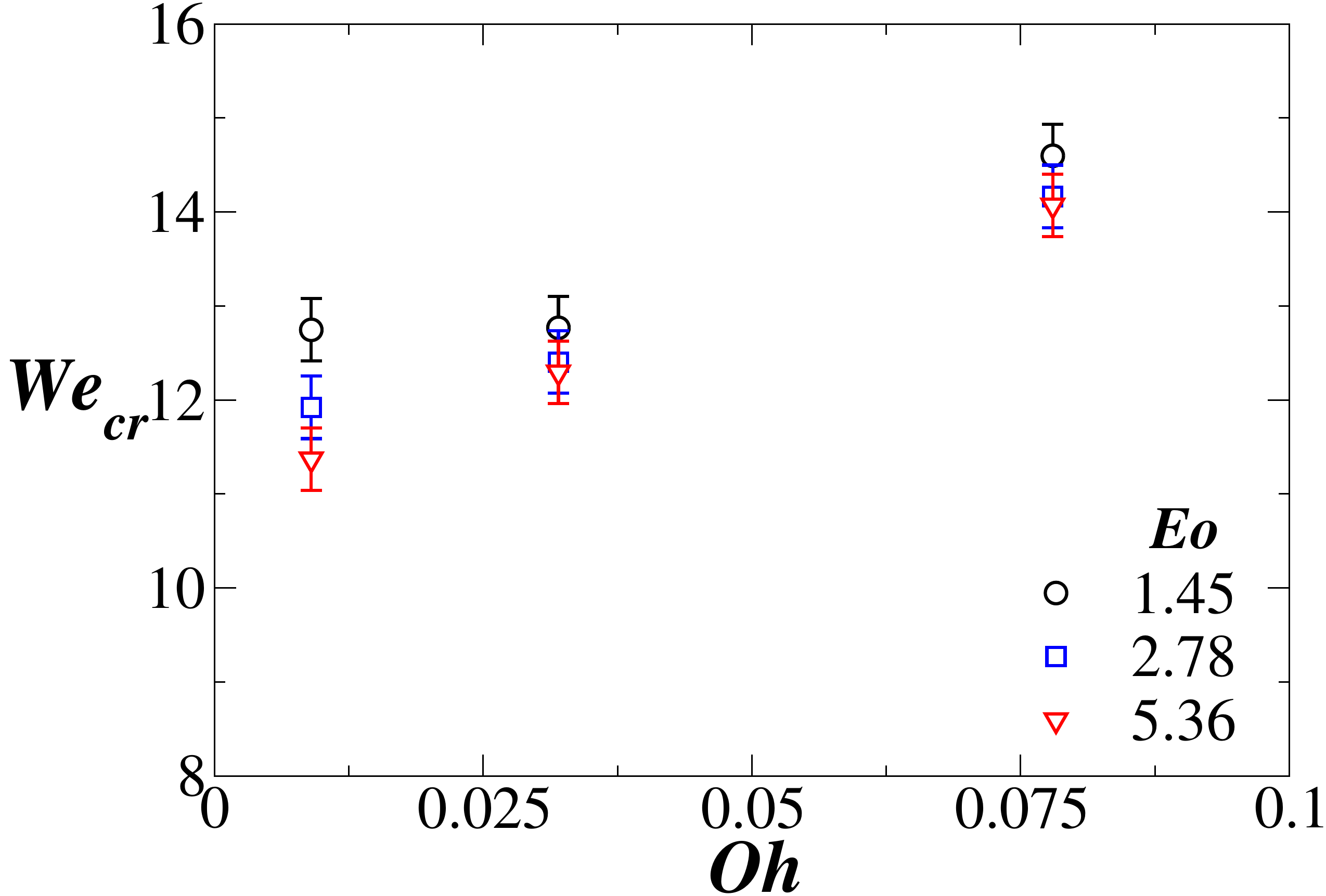} \hspace{2mm} \includegraphics[width=0.45\textwidth]{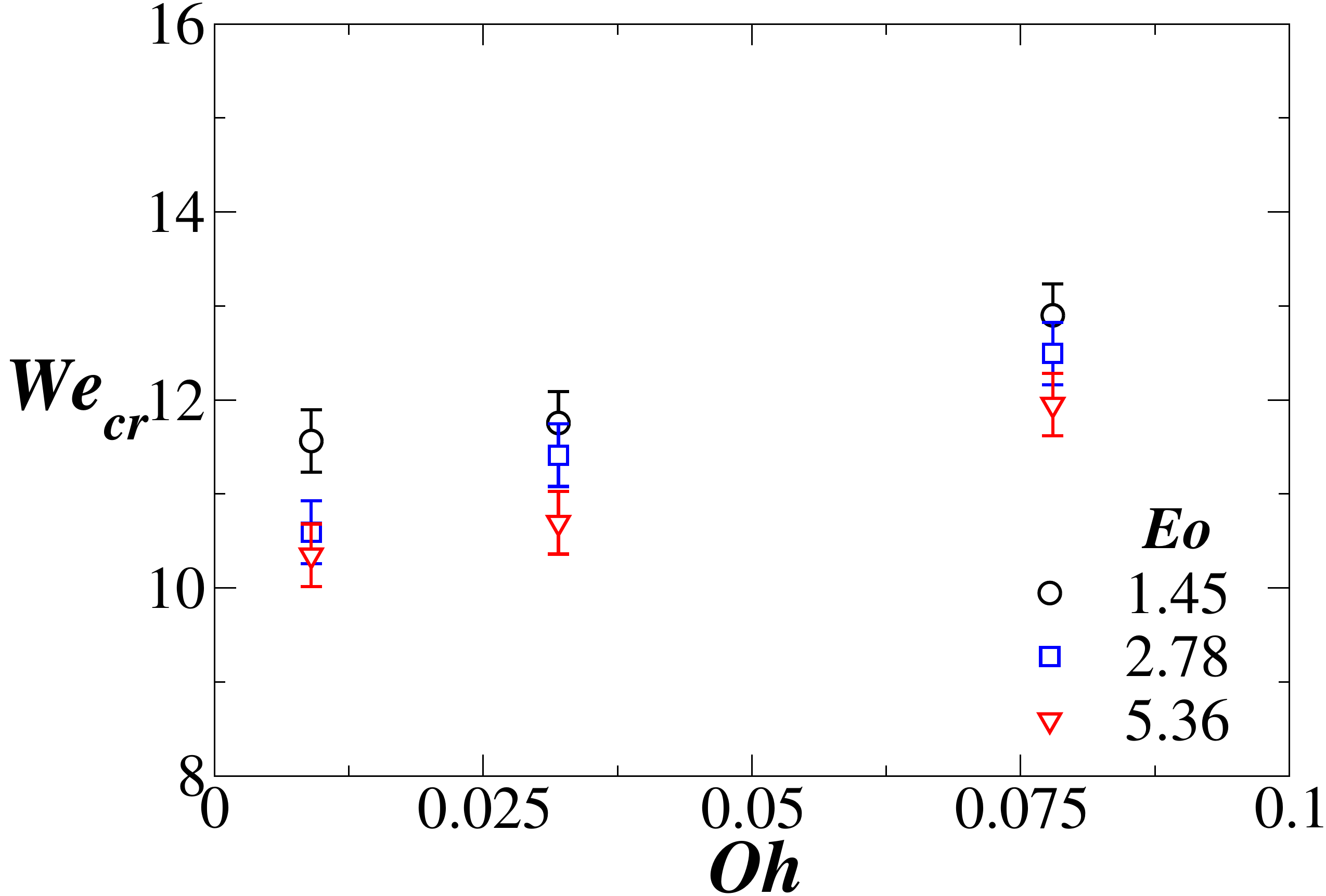} \\
\hspace{0.6cm} {\large (c) $\alpha = 60^\circ$}  \\
\includegraphics[width=0.45\textwidth]{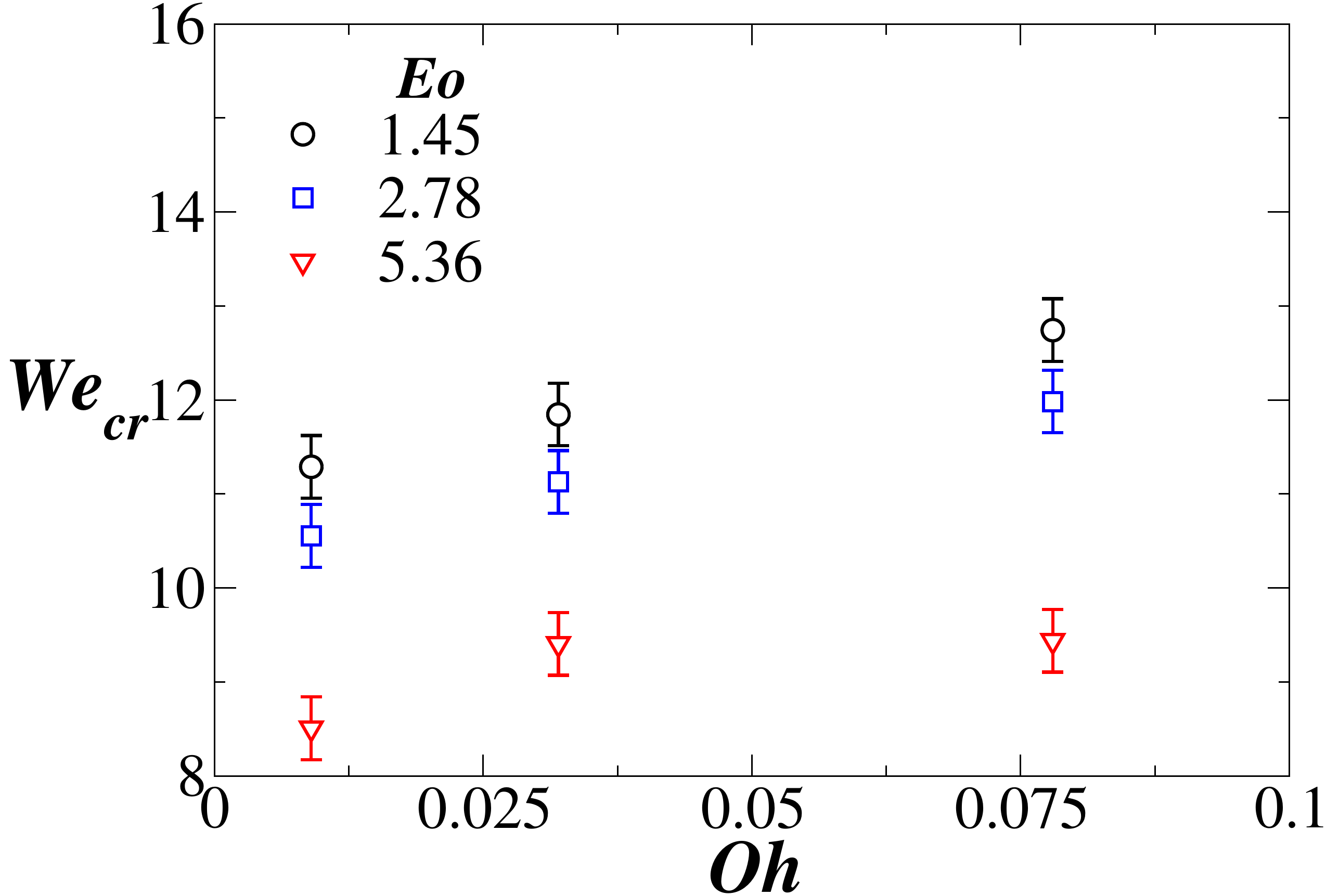}
\caption{The critical Weber number, $We_{cr}$ versus Ohnesorge number, $Oh$ for different values of $Eo$: (a) $\alpha=0^\circ$, (b) $\alpha=30^\circ$ and (c) $\alpha=60^\circ$.}
\label{fig12}
\end{figure}

Two dimensionless numbers, namely the Weber number, $We$ and the Ohnesorge number, $Oh$ are found to be effective in characterising the breakup regimes and transition between the vibrational and the bag breakup regimes. For relatively low Ohnesorge numbers ($Oh < 0.1$), the Weber number has a dominant influence on droplet breakup, and it determines the breakup regimes as discussed above for $Oh\approx0.0015$ (water droplets). In Fig. \ref{fig12}(a-c), we present the variations of $We_{cr}$ versus $Oh$ for different values of $Eo$ and $\alpha$. The value of $Oh$ is varied by changing the concentration of the glycerol-water solution. As expected. it can be seen that $We_{cr}$ increases with increasing $Oh$ for all the cases. Aalburg {\it et al.} \cite{Aalburg2003} also found that the effect of surface tension, which is responsible to oppose the fragmentation, decreases with increasing $Oh$.

Finally, we investigate the variations of $We_{cr}$ with $\alpha$ for different values of $Eo$ and for different glycerol-water solutions (for different values of $Oh$). It can be seen that $We_{cr}$ decreases with the increase in $\alpha$ irrespective of the liquids considered. However, as observed for a low value of $Oh$ (see Fig. \ref{fig7}), $We_{cr}$ did not approach the asymptotic value of $6$ observed in the opposed flow configuration \cite{Villermaux2009}. The Ohnesorge number, $Oh$ has been accounted in the correlation to predict the modified critical Weber number, $We_{cr}^m$ of a viscous droplet as follows \cite{pilch1987,guildenbecher2009,kulkarni2014}
\begin{equation}\label{eq:1}
We_{cr}^m=We_{cr} (1+ c_1 Oh^{c_2}),
\end{equation}
where the values of $c_1$ and $c_2$ are 0.776 and 0.338 respectively, which are obtained using a regression analysis. In the present study, the mean absolute percentage error between the predicted and experimental value is found to be $<6 \%$. 
    
\begin{figure}[H]
\centering
\hspace{1cm} {\large (a) $Oh\approx0.009$} \hspace{6.5cm} {\large (b) $Oh\approx0.032$}\\
\includegraphics[width=0.45\textwidth]{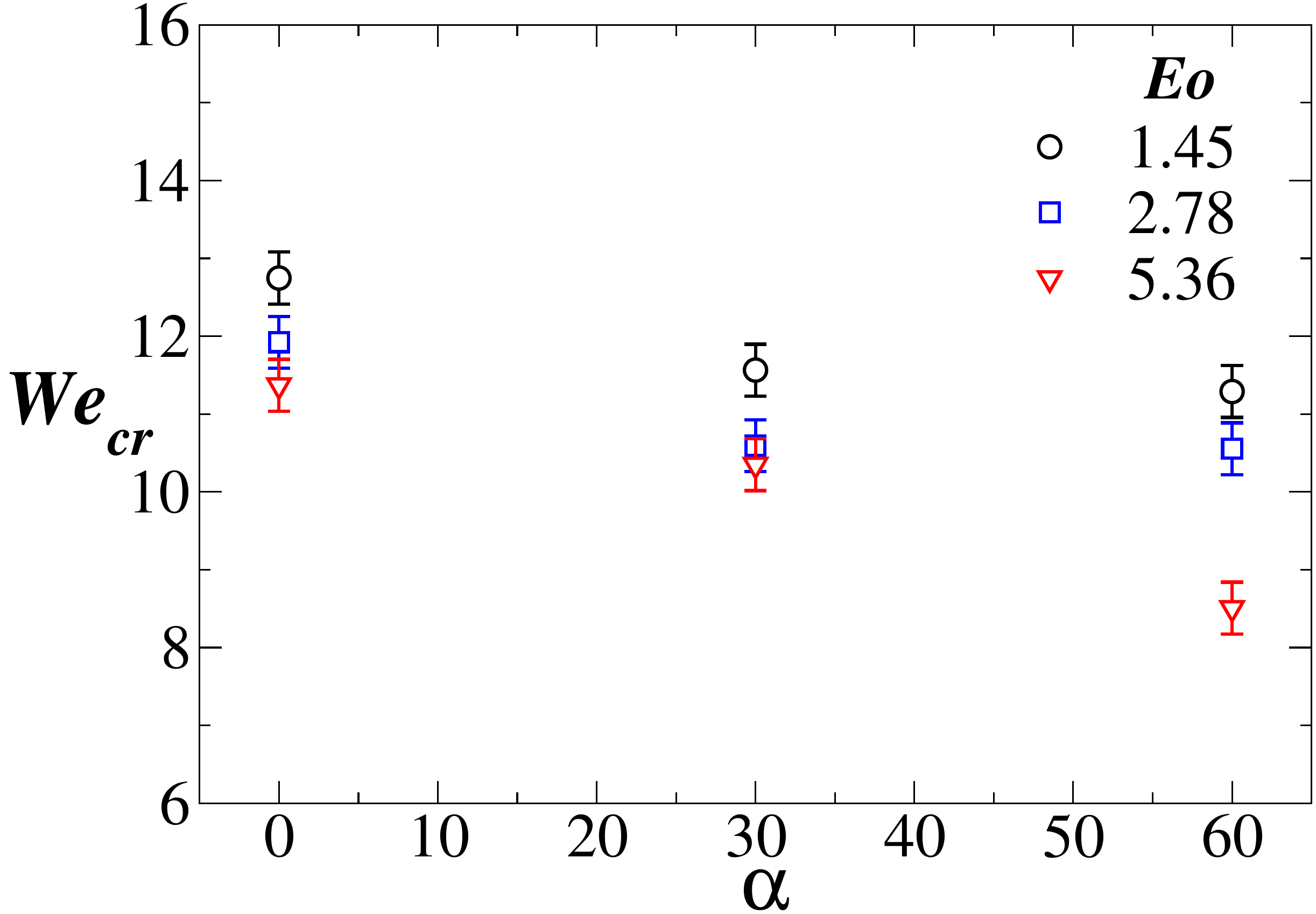} \hspace{2mm} \includegraphics[width=0.45\textwidth]{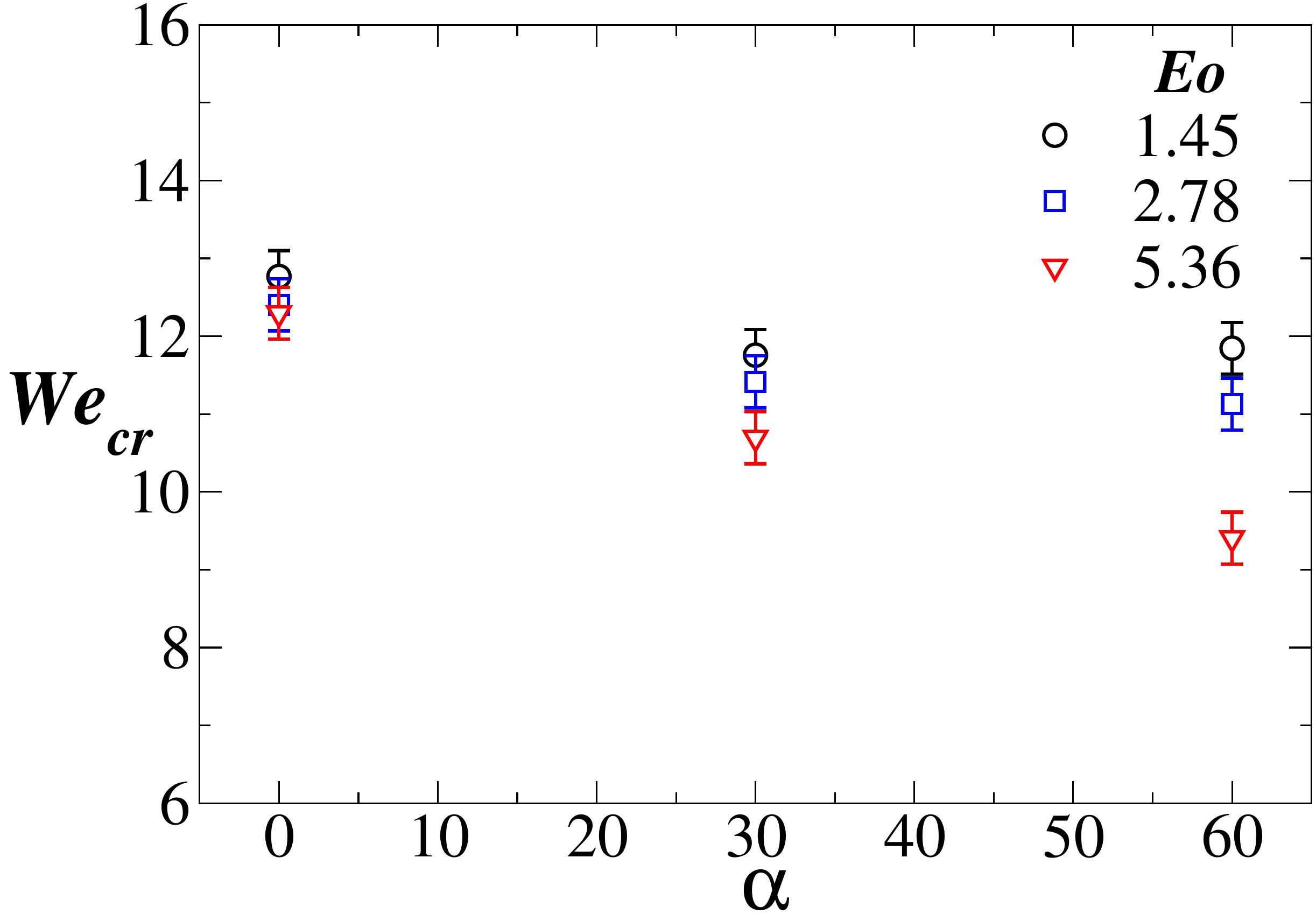} \\
\hspace{0.6cm}  (c) {\large $Oh\approx0.078$}  \\
\includegraphics[width=0.45\textwidth]{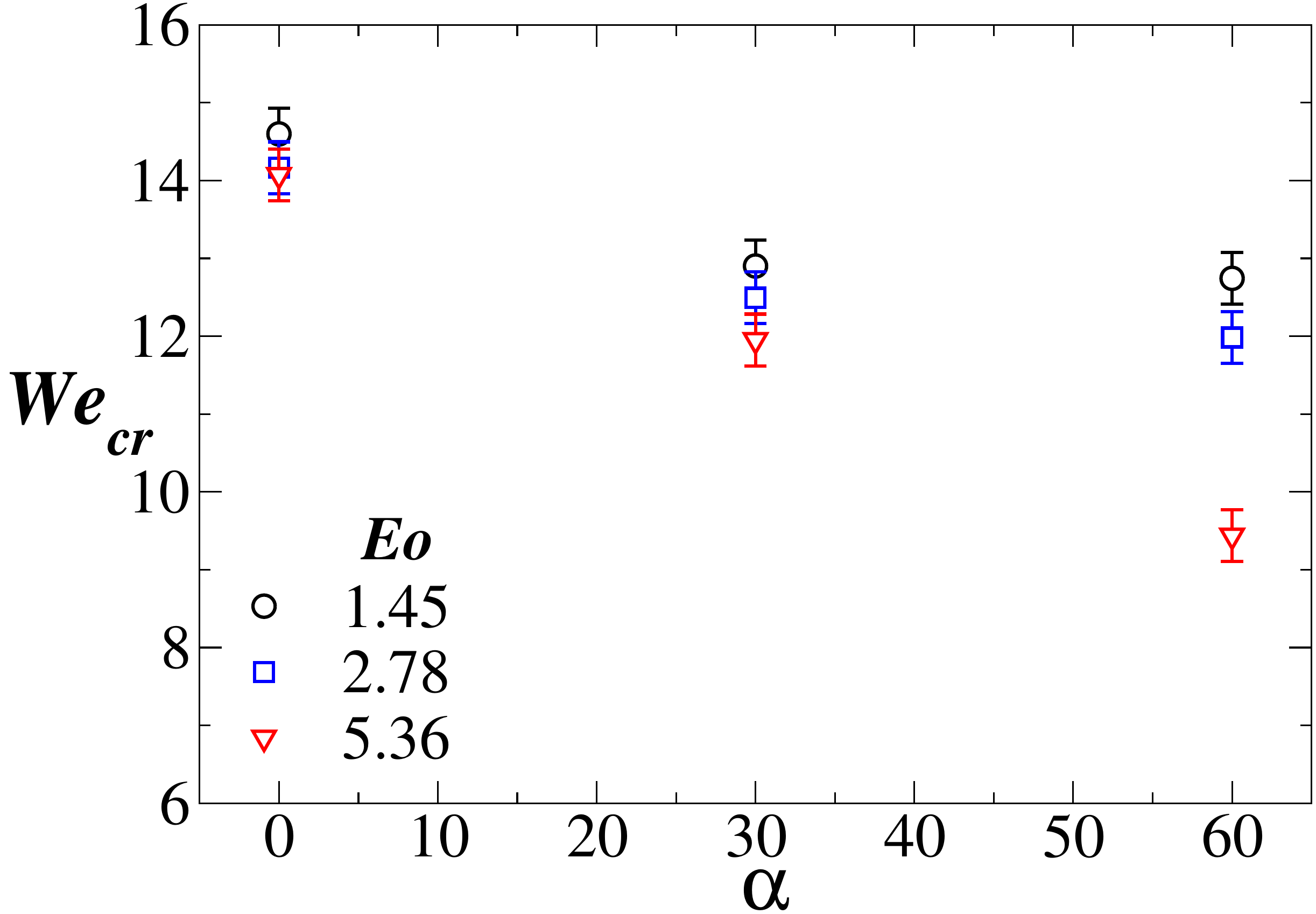}
\caption{Variations of the critical Weber number, $We_{cr}$ versus $\alpha$ (in degree) for (a) 50\% glycerol + 50\% water solution ($Oh \approx 0.009$), (b) 70\% glycerol + 30 \%water solution ($Oh \approx 0.032$) and (c) 80\% glycerol + 20\% water solution ($Oh \approx 0.078$).}
\label{fig13}
\end{figure}

\section{Concluding remarks}
\label{sec:conc}
    \label{sec:conc}
    We have conducted experiments to investigate the effect of the air stream obliquity $(\alpha)$, the droplet diameter $(D)$ and the fluid properties of the liquid on the deformation and the breakup phenomena of droplets experiencing an oblique continuous air stream. A high-speed imaging system is used to record the trajectories and topological changes of the droplets. Four different liquids, namely, deionised water, 50\%, 70\% and 80\% aqueous mixtures of glycerol (by volume) are considered in our experiments. The angle of the continuous air stream is varied from $\alpha=0^{\circ}$ to $\alpha=60^{\circ}$ with an interval of $10^{\circ}$. The values of the critical Weber number $(We_{cr})$ for the bag-type breakup are obtained as a function of the E\"{o}tv\"{o}s number $(Eo)$, angle of inclination of the air stream $(\alpha)$ and the Ohnesorge number $(Oh)$. It is found that, in case of the oblique air stream, the droplet entering the air stream follows a rectilinear motion that transforms into a curvilinear motion when the droplet undergoes topological changes. As we gradually change the cross-flow configuration to the in-line configuration, a sharp decrease in the critical Weber number is observed, which asymptotically reaches to the same value as that observed in an in-line (opposed) flow configuration. The droplet breakup time also increases with the increase in the oblique angle of the continuous air stream. This behaviour can be attributed to the deceleration of the droplet as it follows a curvilinear trajectory and shifts its direction to the nearly in-line (co-flow) arrangement. Thus increasing the residence time in a short path length leads to a more effective energy transfer from the air stream to the droplets, which dissipates to overcome the viscous and the surface tension forces, to form smaller satellite droplets. As expected, increasing the droplet diameter (or increasing $Eo$, which signifies a decrease in the relative influence of the surface tension force over the gravitational force) decreases the value of $We_{cr}$. Increasing the viscosity of the droplet (which is equivalent to increasing the value of $Oh$) requires higher $We$ for the droplet to undergo bag-type breakup.

\section*{Acknowledgments}
Authors would like to acknowledge the financial support from the Department of Science and Technology, India (SERB grant No. ECR/2015/000365). SKS and PKK also thank the Ministry of Human Resource Development, India for providing research fellowships.


\begin{thebibliography}{10}

\bibitem{sikroria2014experimental}
T. Sikroria, A. Kushari, S. Syed, and J.~A. Lovett, ``Experimental
  investigation of liquid jet breakup in a cross flow of a swirling air
  stream,'' J. Eng. Gas Turb. Power {\bf 136},  061501  (2014).

\bibitem{Villermaux2009}
E. Villermaux and B. Bossa, ``Single-drop fragmentation determines size
  distribution of raindrops,'' Nature Physics {\bf 5},  697  (2009).

\bibitem{varga2003initial}
C.~M. Varga, J.~C. Lasheras, and E.~J. Hopfinger, ``Initial breakup of a
  small-diameter liquid jet by a high-speed gas stream,'' J. Fluid Mech. {\bf
  497},  405  (2003).

\bibitem{lefebvre2017atomization}
A.~H. Lefebvre and V.~G. McDonell, {\em Atomization and sprays} (CRC press,
  ADDRESS, 2017).

\bibitem{taylor1963shape}
G.~I. Taylor, ``The shape and acceleration of a drop in a high speed air
  stream,'' The scientific papers of GI Taylor {\bf 3},  457  (1963).

\bibitem{Jain2015}
M. Jain, R.~S. Prakash, G. Tomar, and R.~V. Ravikrishna, ``Secondary breakup of
  a drop at moderate Weber numbers,'' Proc. R. Soc. Lond. A {\bf 471},
  20140930  (2015).

\bibitem{dai2001temporal}
Z. Dai and G.~M. Faeth, ``Temporal properties of secondary drop breakup in the
  multimode breakup regime,'' Int. J. Multiphase Flow {\bf 27},  217  (2001).

\bibitem{pilch1987}
M. Pilch and C.~A. Erdman, ``Use of breakup time data and velocity history data
  to predict the maximum size of stable fragments for acceleration-induced
  breakup of a liquid drop,'' Int. J. Multiphase Flow {\bf 13},  741  (1987).

\bibitem{chou1998temporal}
W.-H. Chou and G.~M. Faeth, ``Temporal properties of secondary drop breakup in
  the bag breakup regime,'' Int. J. Multiphase Flow {\bf 24},  889  (1998).

\bibitem{krzeczkowski1980measurement}
S.~A. Krzeczkowski, ``Measurement of liquid droplet disintegration
  mechanisms,'' Int. J. Multiphase Flow {\bf 6},  227  (1980).

\bibitem{guildenbecher2009}
D.~R. Guildenbecher, C. L{\'o}pez-Rivera, and P.~E. Sojka, ``Secondary
  atomization,'' Exp. Fluids {\bf 46},  371  (2009).

\bibitem{cao2007}
X.-K. Cao {\it et~al.}, ``A new breakup regime of liquid drops identified in a
  continuous and uniform air jet flow,'' Phys. Fluids {\bf 19},  057103
  (2007).

\bibitem{Suryaprakash2019}
S. R. and G. Tomar, ``Secondary atomization,'' J. Indian I. Sci. {\bf 99},  77
  (2019).

\bibitem{kulkarni2014}
V. Kulkarni and P.~E. Sojka, ``Bag breakup of low viscosity drops in the
  presence of a continuous air jet,'' Phys. Fluids {\bf 26},  072103  (2014).

\bibitem{flock2012}
A.~K. Flock {\it et~al.}, ``Experimental statistics of droplet trajectory and
  air flow during aerodynamic fragmentation of liquid drops,'' Int. J.
  Multiphase Flow {\bf 47},  37  (2012).

\bibitem{hanson1963}
A.~R. Hanson, E.~G. Domich, and H.~S. Adams, ``Shock tube investigation of the
  breakup of drops by air blasts,'' Phys. Fluids {\bf 6},  1070  (1963).

\bibitem{wierzba1990deformation}
A. Wierzba, ``Deformation and breakup of liquid drops in a gas stream at nearly
  critical Weber numbers,'' Exp. Fluids {\bf 9},  59  (1990).

\bibitem{krzeczkowski1980}
S.~A. Krzeczkowski, ``Measurement of liquid droplet disintegration
  mechanisms,'' Int. J. Multiphase Flow {\bf 6},  227  (1980).

\bibitem{hsiang1993}
L.~P. Hsiang and G.~M. Faeth, ``Drop properties after secondary breakup,'' Int.
  J. Multiphase Flow {\bf 19},  721  (1993).

\bibitem{wang2014}
C. Wang, S. Chang, H. Wu, and J. Xu, ``Modeling of drop breakup in the bag
  breakup regime,'' Appl. Phys. Lett. {\bf 104},  154107  (2014).

\bibitem{dai2001}
Z. Dai and G.~M. Faeth, ``Temporal properties of secondary drop breakup in the
  multimode breakup regime,'' Int. J. Multiphase Flow {\bf 27},  217  (2001).

\bibitem{chou1998}
W.~H. Chou and G.~M. Faeth, ``Temporal properties of secondary drop breakup in
  the bag breakup regime,'' Int. J. Multiphase Flow {\bf 24},  889  (1998).

\bibitem{zhao2016}
H. Zhao {\it et~al.}, ``Influence of surfactant on the drop bag breakup in a
  continuous air jet stream,'' Phys. Fluids {\bf 28},  054102  (2016).

\bibitem{Inamura2009}
T. Inamura, H. Yanaoka, and T. Kawada, ``Visualization of airflow around a
  single droplet deformed in an airstream,'' Atomization and Sprays {\bf 19},
  667  (2009).

\bibitem{brenner1993}
M. Brenner and A. Bertozzi, ``Spreading of droplets on a solid surface,''
  Physical review letters {\bf 71},  593  (1993).

\bibitem{jalaal2012fragmentation}
M. Jalaal and K. Mehravaran, ``Fragmentation of falling liquid droplets in bag
  breakup mode,'' Int. J. Multiphase Flow {\bf 47},  115  (2012).

\bibitem{nicholls1969}
J.~A. Nicholls and A.~A. Ranger, ``Aerodynamic shattering of liquid drops.,''
  AIAA J. {\bf 7},  285  (1969).

\bibitem{rioboo2002}
R. Rioboo, M. Marengo, and C. Tropea, ``Time evolution of liquid drop impact
  onto solid, dry surfaces,'' Exp. Fluids {\bf 33},  112  (2002).

\bibitem{fp}
``http://www.aciscience.org/docs/physical-properties-of-glycerine-and-its-solutions.pdf,''
   .

\bibitem{merkle2003effect}
K. Merkle, H. Haessler, H. B{\"u}chner, and N. Zarzalis, ``Effect of co-and
  counter-swirl on the isothermal flow-and mixture-field of an airblast
  atomizer nozzle,'' International Journal of Heat and Fluid Flow {\bf 24},
  529  (2003).

\bibitem{gupta2001swirl}
A.~K. Gupta, M.~J. Lewis, and M. Daurer, ``Swirl effects on combustion
  characteristics of premixed flames,'' J. Eng. Gas Turbines Power {\bf 123},
  619  (2001).

\bibitem{Aalburg2003}
C. Aalburg, B.~v. Leer, and G. Faeth, ``Deformation and drag properties of
  round drops subjected to shock-wave disturbances,'' AIAA journal {\bf 41},
  2371  (2003).

\end{thebibliography}

\end{document}